\documentclass[12pt,a4paper]{article}
\usepackage{graphicx}

\def\simgt{\stackrel{>}{{}_\sim}}
\def\simlt{\stackrel{<}{{}_\sim}}

\begin{document}

\pagestyle{empty}

{\Large\bf Supernova 2007bi was a pair-instability explosion}

\vspace{1cm}

\begin{centering}

A.~Gal-Yam,\\
{\scriptsize Benoziyo Center for
Astrophysics, Faculty of Physics,} 
{\scriptsize The Weizmann Institute of Science,
Rehovot 76100, Israel,}\\
P. Mazzali,\\ 
{\scriptsize Max-Planck-Institut f\"{u}r
Astrophysik,} 
{\scriptsize Karl-Schwarzschild-Str. 1, 85748 Garching,
Germany,}\\
{\scriptsize and Scuola Normale Superiore,}
{\scriptsize Piazza Cavalieri 7, 56127 Pisa, Italy},\\
E.~O.~Ofek,\\
{\scriptsize Department of Astronomy, 105-24,}
{\scriptsize California Institute of Technology, Pasadena, CA 91125, USA,}\\
P.~E.~Nugent,\\
{\scriptsize Lawrence Berkeley National Laboratory,} 
{\scriptsize 1 Cyclotron Road, Berkeley, CA 94720-3411, USA,}\\
S.~R.~Kulkarni, M.~M.~Kasliwal, R.~M.~Quimby,\\
{\scriptsize Department of Astronomy, 105-24,}
{\scriptsize California Institute of Technology, Pasadena, CA 91125, USA,}\\
A.~V.~Filippenko, S.~B.~Cenko, R.~Chornock,\\
{\scriptsize Department of Astronomy,}
{\scriptsize University of California, Berkeley, CA 94720-3411, USA,}\\
R.~Waldman,\\
{\scriptsize The Racah Institute of Physics,}
{\scriptsize The Hebrew University, Jerusalem 91904, Israel,}\\
D.~Kasen,\\
{\scriptsize Department of Astronomy and Astrophysics,}
{\scriptsize University of California, Santa Cruz, CA 95064, USA,}\\
M.~Sullivan,\\
{\scriptsize Department of Astrophysics,}
{\scriptsize University of Oxford, Keble Road, Oxford OX1 3RH, UK,}\\
E.~C.~Beshore,\\
{\scriptsize Department of Planetary Sciences, Lunar and Planetary Laboratory,} 
{\scriptsize 1629 E. University Blvd, Tucson AZ 85721, USA,}\\
A.~J.~Drake,\\
{\scriptsize Department of Astronomy, 105-24,}
{\scriptsize California Institute of Technology, Pasadena, CA 91125, USA,}\\
R.~C.~Thomas,\\
{\scriptsize Luis W. Alvarez Fellow,}
{\scriptsize Lawrence Berkeley National Laboratory,}
{\scriptsize 1 Cyclotron Road, Berkeley, CA 94720-3411, USA,}\\
J.~S.~Bloom, D.~Poznanski, A.~A.~Miller,\\
{\scriptsize Department of Astronomy,}
{\scriptsize University of California, Berkeley, CA 94720-3411, USA,}\\
R.~J.~Foley,\\
{\scriptsize Clay Fellow,}
{\scriptsize Harvard-Smithsonian Center for Astrophysics,}
{\scriptsize 60 Garden Street, Cambridge, MA 02138,}\\
J.~M.~Silverman,\\
{\scriptsize Department of Astronomy,}
{\scriptsize University of California, Berkeley, CA 94720-3411, USA,}\\
I.~Arcavi,\\
{\scriptsize Benoziyo Center for
Astrophysics, Faculty of Physics,}
{\scriptsize The Weizmann Institute of Science,
Rehovot 76100, Israel,}\\
R.~S.~Ellis,\\
{\scriptsize Department of Astronomy, 105-24,}
{\scriptsize California Institute of Technology, Pasadena, CA 91125, USA,}\\
J.~Deng,\\
{\scriptsize National Astronomical Observatories,}
{\scriptsize Chinese Academy of Sciences, Beijing 100012, China.}

\end{centering}

\date{\today}{}

\clearpage


{\bf Stars with initial masses
$10$\,M$_{\odot} \simlt$ $M_{\rm initial}\simlt 100$\,M$_{\odot}$
fuse progressively heavier elements in their centres, up to
inert iron. The core then gravitationally collapses to a neutron star 
or a black hole, leading
to an explosion -- an iron-core-collapse supernova (SN)\cite{sma2009,gal+2009}. 
In contrast, extremely massive stars ($M_{\rm initial} \simgt 140\,$M$_{\odot}$), if such exist, have oxygen cores which exceed $M_{\rm core} = 50$\,M$_{\odot}$.
There, high temperatures are reached at relatively low densities. Conversion
of energetic, pressure-supporting photons into electron-positron pairs
occurs prior to oxygen ignition, and leads to
a violent contraction that triggers a catastrophic nuclear
explosion\cite{rak+1967,bar+1967,heg+2002}.
Tremendous energies ($\simgt 10^{52}$\,erg) are released, 
completely unbinding the star in a pair-instability SN (PISN),
with no compact remnant. Transitional objects with
$100$\,M$_{\odot} < M_{\rm initial} < 140$\,M$_{\odot}$,
which end up as iron-core-collapse supernovae following
violent mass ejections, perhaps due to
short instances of the pair instability, may have been 
identified\cite{wos+2007,smi+2009,mil+2009b}.
However, genuine PISNe, perhaps common
in the early Universe, have not been observed to date.
Here, we present our discovery of
SN 2007bi, a luminous, slowly evolving supernova located
within a dwarf galaxy ($\sim 1\%$ the size of the Milky Way). We
measure the exploding core mass to be likely
$\sim 100$\,M$_{\odot}$, in which case theory unambiguously 
predicts a PISN outcome. 
We show that $>3$\,M$_{\odot}$ of radioactive
$^{56}$Ni were synthesized, and that
our observations are well fit
by PISN models\cite{kas+2008,wal2008}. A PISN explosion
in the local Universe indicates
that nearby dwarf galaxies probably host extremely massive stars, above the
apparent Galactic limit\cite{fig2005}, perhaps resulting from star
formation processes similar to those that created the first stars in
the Universe.
}

\clearpage



We discovered a new optical transient (SNF20070406-008)
on 2007 April 6.5 (UT dates are used throughout this paper) at right ascension
$\alpha = 13^h19^m20.2^s$ and declination $\delta = 08^{\circ}55'44.0''$ (J2000). 
Follow-up spectroscopic
observations showed that this was a supernova (SN 2007bi),
with no trace of either hydrogen or helium, leading to
a Type Ic classification\cite{fil97}, albeit of a peculiar nature with
only one previously known counterpart (SN 1999as\cite{kas2004}; Fig. 1).
No signs of interaction with circumstellar material (CSM, a major
source of uncertainty in the analysis of previous 
luminous SNe\cite{agn+2009,smi+2009})
are observed throughout the evolution of this event (Fig. 1).
A search for pre-explosion data recovered observations by the
Catalina Sky Survey (CSS\cite{dra+2009})
that allowed an accurate determination of
the date and magnitude of the supernova brightness peak. Photometric
observations continued for 555 days during our intense follow-up
campaign. A red ($R$-band) light curve is plotted in Fig. 2 (top).

The measured light curve is unique, showing a very long rise time to peak 
($\sim70$\,days; Fig. 2; SI $\S~2$),
an extreme luminosity reaching an absolute peak $R$-band magnitude of 
$M_R = -21.3$\,mag, and a slow decline
($0.01$\,mag\,day$^{-1}$ over $>500$ days), consistent with the decay rate 
of radioactive $^{56}$Co. These properties suggest that the very massive
ejecta were energized by a large amount of radioactive nickel 
($>3$\,M$_{\odot}$; Fig. 2, 3; SI $\S~3$),
as expected from pair-instability SN models\cite{bar+1967,heg+2002,wal2008}.
Our spectra, lacking any signs of hydrogen
or helium, indicate that this mass is dominated by C, O, and heavier elements.
The large amount of kinetic energy released, $E_{\rm k} \approx 10^{53}$\,erg
(Fig. 2; SI $\S~3$), is
comparable to those derived for the most energetic gamma-ray bursts
(GRBs\cite{cen+2009}), placing this event among the most extreme explosions known.
In Fig. 2 (bottom) we show theoretical light curves calculated
from PISN models\cite{heg+2002,kas+2008} prior to our discovery.
The data fit the models very well, suggesting we observed the explosion of
a star with a helium core mass around $100$\,M$_{\odot}$.

PISN models imply that such an explosion would synthesize 3--10\,M$_{\odot}$ of
radioactive $^{56}$Ni (Table 1). Such a large amount of newly synthesized radioactive
material would energize the SN debris for an extended period of time,
ionizing the expanding gas cloud. Collisional excitation would lead to strong nebular
emission lines, whose strength should be roughly proportional to the amount
of radioactive source material, providing another testable prediction.
Fig. 3 (top) shows a comparison of the nebular
spectrum of SN 2007bi with that of the well-studied, $^{56}$Ni-rich SN 1998bw,
which produced $\sim 0.5$\,M$_{\odot}$ of radioactive nickel\cite{maz+2001},
suggesting that SN 2007bi produced $\simgt 7$\,M$_{\odot}$ of nickel 
(SI $\S~3$), again supporting the PISN interpretation.

Modelling the nebular spectrum, we are able to resolve the elemental composition
of the fraction of the ejected mass that is illuminated by radioactive nickel. 
We can directly measure the abundances
of C, O, Na, Mg, Ca, and Fe, and derive the mass of radioactive $^{56}$Ni. Our elemental
abundance ratios are in good agreement with model predictions\cite{heg+2002}
for heavier elements, while lighter elements (C, O, Mg) seem to be underobserved.
Adopting the calculated model output\cite{heg+2002} for elements which do not 
have strong nebular emission in the optical (mostly Si and S, and some Ne and Ar), 
we arrive at a total illuminated mass of
$>50$\,M$_{\odot}$, with a composition as described in Table 1. Note that this falls
well below the total mass derived from the photometry, indicating that even the
unprecedented amount of radioactive nickel produced by SN 2007bi was not sufficient
to energize the entire mass ejected by this extreme explosion (SI $\S~6$).
The unilluminated mass
probably contains more light elements that originated in the outer envelopes of the
exploding star, and seem deficient in our nebular observations (see SI $\S~3,6$ for
additional details).

Our data thus provide strong evidence that we have observed the explosion of a 
helium core with $M \approx$ 100\,M$_{\odot}$, which, according to 
theory, can only result in a PISN\cite{rak+1967,bar+1967,wos+2007,heg+2002,wal2008,lan+2007}. 
The measured light curve, radioactive nickel yield, and elemental composition of the ejecta are
consistent with models of PISNe that were calculated before our discovery. 
Based on fewer observations of SN 2007bi, combined with their analysis of 
the host-galaxy properties, ref. \cite{you+2009} consider both a PISN model and a massive
iron-core-collapse SN interpretation\cite{ume+2008}, slightly favoring the latter.
However, our quantitative estimate of the
helium core mass from our peak light-curve shape and the analysis of the
nebular spectra, is inconsistent with iron-core-collapse
models\cite{ume+2008} and theoretically requires a PISN\cite{heg+2002,wal2008}. 
We thus conclude that we have most likely discovered the first clear example of a PISN.

There are several implications of this discovery. Theory allows stars as massive as
1000\,M$_{\odot}$ to have formed in the very early Universe\cite{bro+2004}. However,
the most massive stars known in the local Universe (e.g.,
luminous blue variables) have estimated masses around 
150\,M$_{\odot}$\cite{fig2005}. In the single
example known so far, such a hypergiant star exploded in a regular core-collapse event
(SN 2005gl)\cite{gal+2009}. Our detection of a PISN from a $\sim 100$\,M$_{\odot}$
core suggests a progenitor with an estimated initial mass around 
200\,M$_{\odot}$\cite{heg+2002}, 
assuming very low mass-loss rates appropriate
for zero-metallicity stars. We note that
this estimate is highly sensitive to poorly understood mass loss, and that
high-metallicity mass-loss prescriptions would require an even higher initial mass.
In a sense, our discovery of such a core is in conflict with commonly used mass-loss 
calculations, which do not allow such a high-mass core to form at the 
measured metallicities\cite{you+2009}. Our finding probably requires the 
modification of mass-loss paradigms, perhaps through increased clumping in massive star 
winds\cite{lan+2007,agn+2009,smi+2009}, at least during the hydrogen-rich mass-loss phase. 

Regardless of the exact mass loss adopted,
our data indicate that extremely massive stars above the Galactic limit
($M > 150$\,M$_{\odot}$) are formed in the local Universe. Perhaps the dwarf,
metal-poor host galaxy of SN 2007bi ($M_B=-16.3$ mag at $z = 0.1279$, indicating an
approximate metallicity of $12+$log$[{\rm O}/{\rm H}]=8.25$ [ref. \cite{tre+2004}];
see ref. \cite{you+2009} for additional details)
retained conditions that were similar to those prevalent in the early
Universe. Luminous events like SN 2007bi can therefore serve as beacons, focusing
our attention on otherwise unremarkable local dwarf galaxies that can be used as fossil
laboratories to study the early Universe. 

Our observational confirmation of PISN
models supports their use to predict the detectability and observed properties of
PISNe from the first stars by future missions such as the James Webb Space Telescope,
to estimate their contribution to the chemical evolution of the Universe\cite{heg+2002},
and to calculate their impact on the
re-ionization of the Universe. With the advent of new wide-field
surveys such as the Palomar Transient Factory\cite{law+2009,rau+2009}
and CRTS\cite{dra+2009}
that monitor millions of nearby low-luminosity,
anonymous galaxies\cite{you+2008},
many additional such events should soon be discovered and 
will further illuminate these important questions.





~\\
We gratefully acknowledge advice and help from E. Pian,
and discussions with Z. Barkat, E. Livne, E. Nakar, N. Langer, 
and P. Podsiadlowski, as well as helpful suggestions and corrections 
by our referees. This work benefited from useful interaction
during the Fireworks meetings held at the Weizmann Institute (2008)
and at the University of Bonn (2009). 
Work related to the CSS data reported here was
supported by the National Aeronautics
and Space Administration (NASA) under a grant issued through the
Science Mission Directorate Near-Earth Object Observations program.
The joint work of A.G. and P.A.M. is supported
by a Weizmann-Minerva grant.
A.G. acknowledges support by the Israeli Science Foundation, an
EU Seventh Framework Programme Marie Curie IRG fellowship, and the
Benoziyo Center for Astrophysics, 
a research grant from the Peter and Patricia Gruber Awards,
and the William Z. and Eda Bess Novick New Scientists Fund at the
Weizmann Institute. A.V.F.'s group at UC Berkeley is grateful for
financial support from the US National Science Foundation, 
the US Department of Energy, the TABASGO Foundation, 
Gary and Cynthia Bengier, and 
the Richard and Rhoda Goldman Fund. 
J.D is supported by the NSFC and by the Chinese 973 Program.
Based in part on data from the W. M. Keck Observatory, which is operated
as a scientific partnership among the California Institute of Technology,
the University of California, and NASA; it was made possible by the
generous financial support of the W. M. Keck Foundation.
This work made use of the NASA/IPAC Extragalactic Database (NED)
which is operated by the Jet Propulsion Laboratory,
California Institute of Technology, under contract with NASA.

%


~\\
A.G. initiated, coordinated, and managed the project, carried out photometric
and spectroscopic analysis, and wrote the manuscript. P.A.M. was responsible to
obtaining the VLT late-time observations, carried out
spectroscopic modelling, and led the theoretical interpretation effort.  
E.O.O. led the Palomar photometry effort, obtained P200 and Keck observations,
and performed the photometric calibration analysis. P.E.N. discovered SN 2007bi, 
identified its peculiarity and similarity to SN 1999as, initiated some of the 
early spectroscopic analysis, and led the recovery of pre-discovery data from
DeepSky and the CRTS. S.R.K., M.M.K., and R.M.Q. obtained key late-time Keck spectra
and helped with the P60 observations. A.V.F., S.B.C., and R.C.
analyzed early Keck data, and contributed to manuscript preparation
and editing, including final proofreading (A.V.F.). R.W. and D.K. carried out custom 
PISN modelling to compare with the observations. M.S. undertook custom reduction 
of the key late-time Keck spectrum. E.C.B. is the PI of CSS, and his team
acquired the CSS data and provided preliminary calibration of the
results. A.J.D. helped recover CSS data and advised about
their calibration. R.C.T. analyzed early spectra using his automated SYNOW code. 
J.S.B., D.P., and A.A.M. obtained early spectroscopic observations of SN 2007bi 
as well as IR observations with PAIRITEL,
and contributed to analysis and manuscript editing. R.J.F. and J.M.S. contributed
to spectral observations and reductions and advised during manuscript preparation.  
I.A. helped with P60 photometry and calibration and with manuscript editing. 
R.S.E. obtained Keck observations of SN 2007bi. J. Deng contributed to the VLT 
program that resulted in the observations of SN 2007bi and proofread the
manuscript.
%

~\\
The authors declare no competing financial interests.\\

\noindent Correspondence should be addressed to A. Gal-Yam (avishay.gal-yam@weizmann.ac.il.)\\

\noindent Supplementary Information accompanies the paper on www.nature.com/nature\\


\clearpage

\begin{table}
\begin{scriptsize}
\begin{tabular}{|c|c|c|c|c|c|c|c|c|c|c|c|}
\hline
Element [M$_{\odot}$]& C & O & Ne & Na & Mg & Si & S & Ar & Ca & $^{56}$Ni & Total\tabularnewline
\hline
\hline
Measured (1)& 1.0$^{*}$ & 10.1$^{*}$ & 4.0 & 0.0012$^{*}$ & 0.068$^{*}$ & 22.0 & 10.0 & 1.3 & 0.75$^{*}$ & 4.5$^{*}$ & 53.8\tabularnewline
Measured (2)& 1.0$^{*}$ & 12.0$^{*}$ & 4.0 & 0.0013$^{*}$ & 0.095$^{*}$ & 22.0 & 10.0 & 1.3 & 0.90$^{*}$ & 5.7$^{*}$ & 57.1\tabularnewline
Measured (3)& 1.2$^{*}$ & 14.6$^{*}$ & 4.0 & 0.0018$^{*}$ & 0.13$^{*}$ & 22.0 & 10.0 & 1.3 & 1.00$^{*}$ & 7.4$^{*}$ & 61.7\tabularnewline
Measured (4)& 1.0$^{*}$ & 7.5$^{*}$ & 4.0 & 0.0015$^{*}$ & 0.065$^{*}$ & 22.0 & 10.0 & 1.3 & 0.95$^{*}$ & 3.65$^{*}$ & 50.6\tabularnewline
Measured (5)& 1.0$^{*}$ & 9.1$^{*}$ & 4.0 & 0.0018$^{*}$ & 0.085$^{*}$ & 22.0 & 10.0 & 1.3 & 1.10$^{*}$ & 4.6$^{*}$ & 53.3\tabularnewline
Measured (6)& 1.0$^{*}$ & 11.3$^{*}$ & 4.0 & 0.0023$^{*}$ & 0.12$^{*}$ & 22.0 & 10.0 & 1.3 & 1.00$^{*}$ & 6.0$^{*}$ & 56.8\tabularnewline
\hline
95 M$_{\odot}$ model\cite{heg+2002}& 4.1 & 45.2 & 4.0 & 0.0028 & 4.38 & 21.3 & 8.8 & 1.3 & 0.99 & 2.98 & 95\tabularnewline
100 M$_{\odot}$ model\cite{heg+2002}& 4.0 & 43.9 & 4.1 & 0.0028 & 4.41 & 23.1 & 10.0 & 1.5 & 1.22 & 5.82 & 100\tabularnewline
105 M$_{\odot}$ model\cite{heg+2002}& 3.9 & 42.7 & 3.9 & 0.0028 & 4.40 & 24.5 & 10.8 & 1.7 & 1.40 & 9.55 & 105\tabularnewline
\hline
95 M$_{\odot}$ model\cite{wal2008}& 0.28 & 38.1 & 1.60 & 0.0011 & 2.35 & 20.9 & 13.5 & 2.32 & 1.98 & 5.18 & 95\tabularnewline
100 M$_{\odot}$ model\cite{wal2008}& 0.25 & 36.4 & 1.51 & 0.0010 & 2.45 & 22.6 & 15.0 & 2.54 & 2.23 & 8.00 & 100\tabularnewline
105 M$_{\odot}$ model\cite{wal2008}& 0.23 & 35.2 & 1.47 & 0.0009 & 2.47 & 23.9 & 15.9 & 2.73 & 2.46 & 11.89 & 105\tabularnewline
\hline
\hline
\end{tabular}
\end{scriptsize}
\caption{Predicted PISN ejecta composition compared with our measurements. 
We report six different estimates based on the two available late-time 
spectra (Fig. 3): VLT day 414 ([1]-[3]),
and Keck day 530 ([4]-[6]). Each spectrum is
modeled assuming three different explosion dates: 
45\,days ([1], [4]), 77\,days ([2], [5]), and 113\,days ([3], [6])
prior to peak (observed).
The results are qualitatively similar, with $^{56}$Ni masses of 3.7--7.4\,M$_{\odot}$ 
and total masses 51--62\,M$_{\odot}$.
Elements noted
by an * mark  are directly constrained from optical nebular emission lines. 
Abundances of other elements
are probed through their cooling effects via lines outside of the optical range 
(which are modelled), but
we consider these constraints to be weaker.}
\end{table}


\clearpage

\noindent
{\bf Figure 1}:
{%

Spectra of the unusual Type Ic SN 2007bi. We observed SN 2007bi on 2007 April 15.6 
and 16.4, using LRIS\cite{oke+1995} mounted on the Keck~I 10\,m
telescope. Narrow emission lines (see Fig. 3 for details) indicate a
redshift of $z=0.1279$. A survey of our databases of SN spectra shows that
this event is similar only to a single previous example, SN 1999as\cite{kas2004},
until now the most luminous known SN~Ic by a wide margin. SN 2007bi has a comparable
luminosity (Fig. 2). We identify the most prominent features (marked)
as arising from calcium, magnesium, and iron, and
derive a photospheric velocity of $12,000$ km s$^{-1}$. A model fit (SI
$\S~5$) confirms these line
identifications and shows that the absorption in the blue-ultraviolet part of the
spectrum is dominated by blends of iron, cobalt, and nickel lines. The shallow,
poorly defined trough seen around 5500--6000~\AA\ could arise from blends of
silicon and sulfur lines, while lines of neutral oxygen and sodium that
are usually prominent in SN~Ic spectra are remarkably weak. No hydrogen or
helium lines are seen, confirming the Type Ic classification, and strongly
disfavoring the possibility that CSM interaction contributes significantly to the
remarkable luminosity\cite{agn+2009,smi+2009,mil+2009a}. No narrow
sodium absorption is seen, indicating little absorption by dust in the host
galaxy\cite{you+2009}.
The strong emission line around $7300$~\AA\ seems to arise from [Ca~II]
emission at zero expansion velocity, and is usually observed in SNe only in
late-time nebular spectra.
A complete analysis of our full set of photospheric spectra will be
presented in a forthcoming publication; 
see also additional details in ref. \cite{you+2009}.
}

\clearpage

\begin{figure}
\centerline{\includegraphics[width=6.5in,angle=0]{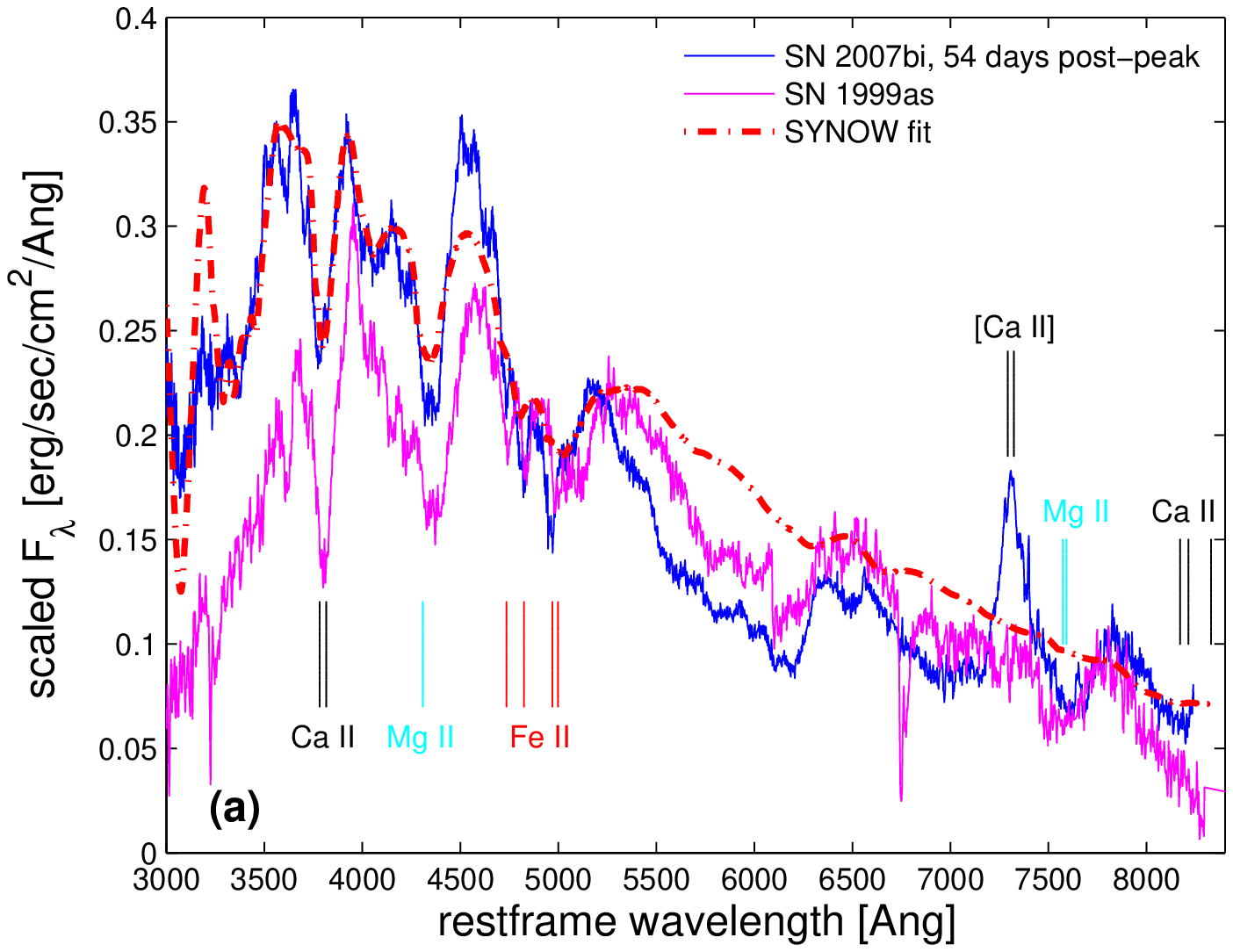}}
\bigskip
\caption[]{~\label{3panelfig}~}
\end{figure}

\clearpage

\noindent
{\bf Figure 2}:
{%

Radioactive $^{56}$Ni and total ejected mass from the light-curve
evolution of SN 2007bi are well fit by PISN models.
{\bf (a)} The $R$-band light curve of SN 2007bi. Data with standard
deviation errors 
are shown. We have compiled
observations obtained by the 48-inch (1.2\,m) Oschin Schmidt, the 
60-inch (1.5\,m) robotic, and the 200-inch (5\,m) Hale telescopes at Palomar Observatory, as
well as photometry from the Catalina Sky Survey (CSS\cite{dra+2009})
and synthetic photometry integrated from our late-time Keck
spectrum (Fig. 3; see SI $\S~1,2$ for additional details).
We find a peak magnitude of $M_R=-21.3\pm0.1$\,mag on 2007 Feb. 21 (SI $\S~2$).
Our error is dominated by the absolute zero-point calibration uncertainty.
The outstanding peak luminosity of this event, if radioactively driven,
suggests that a remarkable amount of $^{56}$Ni was produced ($>3\,$M$_{\odot}$;
ref. \cite{per+2009}; SI $\S~3$).
The slow rise time derived from our fit (77\,days; SI $\S~2$), 
combined with the measured photospheric velocity (12,000\,km\,s$^{-1}$; Fig. 1), 
requires very massive ejecta ($M_{\rm ej} \approx 100$\,M$_{\odot}$)
and a huge kinetic energy release ($E_{\rm k}\approx 10^{53}$\,erg; SI $\S~3$),
adopting the commonly used scaling relations\cite{per+2009,fol+2009}. 
An independent direct estimate for the $^{56}$Ni yield is obtained from the luminosity
during the late-time decay phase, compared to that of SN 1987A\cite{pun+1995} (SI $\S~3$). 
Given the uncertainty in the explosion date of
SN 2007bi and a range of bolometric correction values (SI $\S~2$),
the $^{56}$Ni mass produced by SN 2007bi
was $4$\,M$_{\odot} < M_{^{56}{\rm Ni}} < 7$\,M$_{\odot}$.
The total radiated energy we measure by direct integration of the light curve
is $E_{\rm rad} \approx (1-2) \times 10^{51}$ erg (SI $\S~3$), comparable to
that of the most luminous SNe known\cite{smi+2009}.\\
{\bf (b)} Comparison of the observations of SN 2007bi with models calculated
before the SN discovery\cite{heg+2002,kas+2008}. The curves presented are for various
helium cores (masses as indicated) exploding as PISNe, and cover the
photospheric phase. The data are well fit by 100--110\,M$_{\odot}$ models.
At later times, the emission is nebular and bolometric corrections used to calculate
the model $R$-band light curve cease to hold (SI $\S~4)$. In comparing with 
these restframe models, cosmological time dilation for $z=0.1279$ has been 
taken into account.

\clearpage

\begin{figure}
\centerline{\includegraphics[width=6in,angle=0]{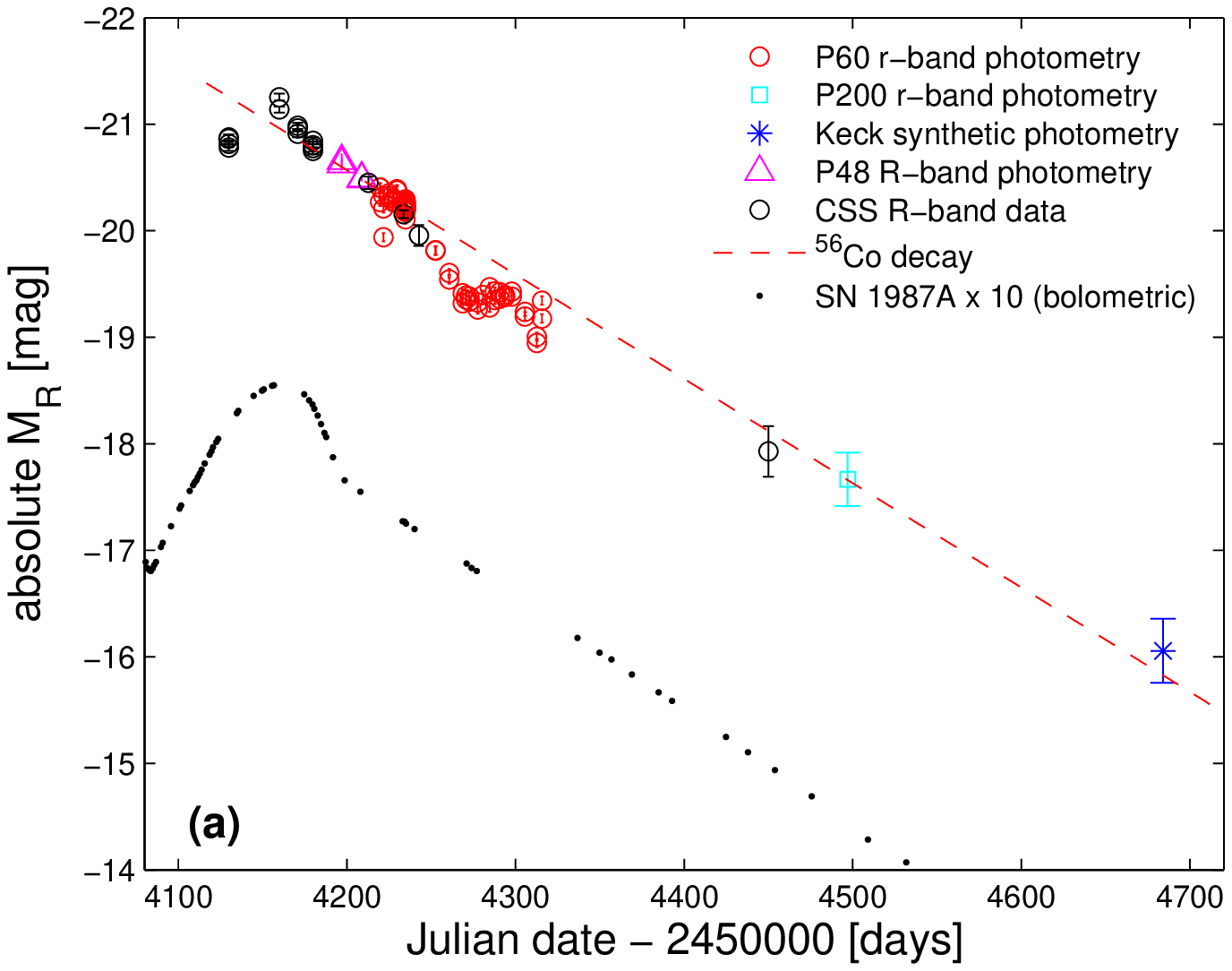}}
\centerline{\includegraphics[width=6in,angle=0]{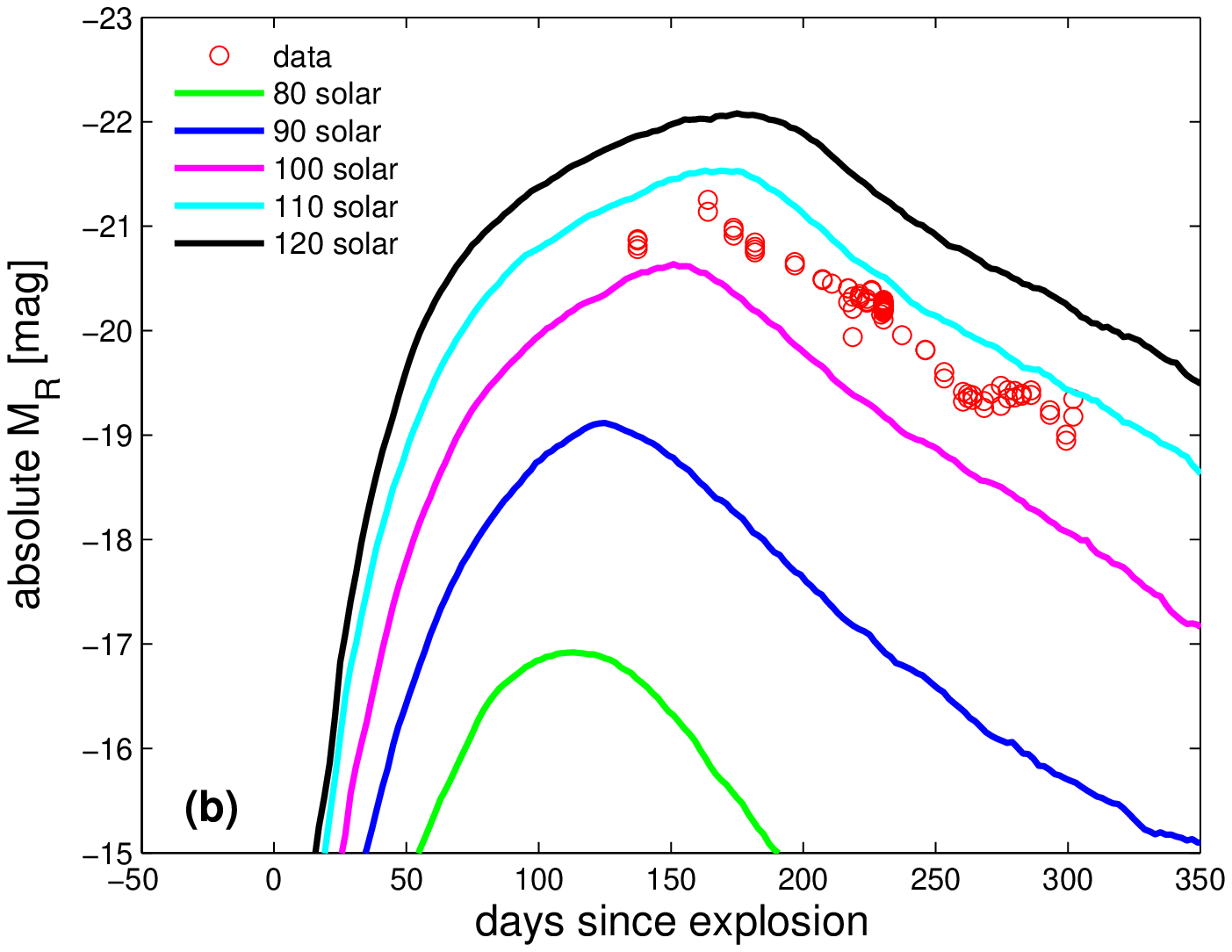}}
\bigskip
\caption[]{~\label{specevolfig}~}
\end{figure}

\clearpage
 
\noindent
{\bf Figure 3}: 

Ejecta composition from nebular spectra of SN 2007bi.
{\bf (a)} Shown are two late-time spectra of SN 2007bi. A spectrum obtained
with the FORS2 spectrograph mounted on the ESO 8.1\,m Very Large Telescope
at Paranal Observatory on 2008 April 10 (414 days post-peak; 367\,days restframe) is not
completely nebular yet. Prominent broad SN emission peaks from neutral and singly ionized
elements are marked. Multiple narrow host-galaxy emission lines ([O~II] $\lambda$3727,
H$\beta$, [O~III] $\lambda\lambda$4959, 5007, H$\alpha$, [S~II] $\lambda\lambda$6716, 6731) 
are seen at $z=0.1279$.
An even later spectrum (530 days post-peak; 470\,days restframe) was obtained with LRIS mounted
on the 10\,m Keck~I telescope on 2008 August 4. This spectrum is fully
nebular, though of lower signal-to-noise ratio.
Comparison with a late-time spectrum of the $^{56}$Ni-rich SN 1998bw,
which produced 0.5\,M$_{\odot}$ of $^{56}$Ni\cite{maz+2001},
adjusted for the larger distance and later restframe spectroscopic 
observations of SN 2007bi and multiplied by eight,
provides a good fit for intermediate-mass elements (O, Na, Mg, Ca) but underpredicts
the strength of the Fe lines. Assuming that emission-line luminosity scales
with the mass of energizing $^{56}$Ni, we derive from the scaling factor
a $^{56}$Ni mass $7.7$\,M$_{\odot} < M_{^{56}{\rm Ni}} < 11.3$\,M$_{\odot}$ (SI $\S~3$),
in reasonable agreement with estimates from early (peak) and late-time (radioactive tail)
photometric estimates. We note that the lack of H$\alpha$ emission at these
late epochs is an especially strong argument against CSM interaction. A
lower limit of $\sim 5 \times 10^{16}$\,cm on the distance
to any H-rich material (in particular, recent mass loss) is derived from
this non-detection assuming an expansion velocity of 12,000\,km\,s$^{-1}$ (Fig. 1).\\
{\bf (b)} Modelling the nebular spectrum. Employing our nebular
spectroscopy code\cite{maz+2001} we are able to constrain the composition of
the ejecta. We find that the emission-line luminosity requires an initial $^{56}$Ni
mass of 3.7--7.4\,M$_{\odot}$ and the composition given in Table 1. As can be
seen there, our measurements are well fit by the theoretical predictions of
PISN models\cite{heg+2002} calculated well before we discovered SN 2007bi.

{%
\clearpage

\begin{figure}
\centerline{\includegraphics[width=6in,angle=0]{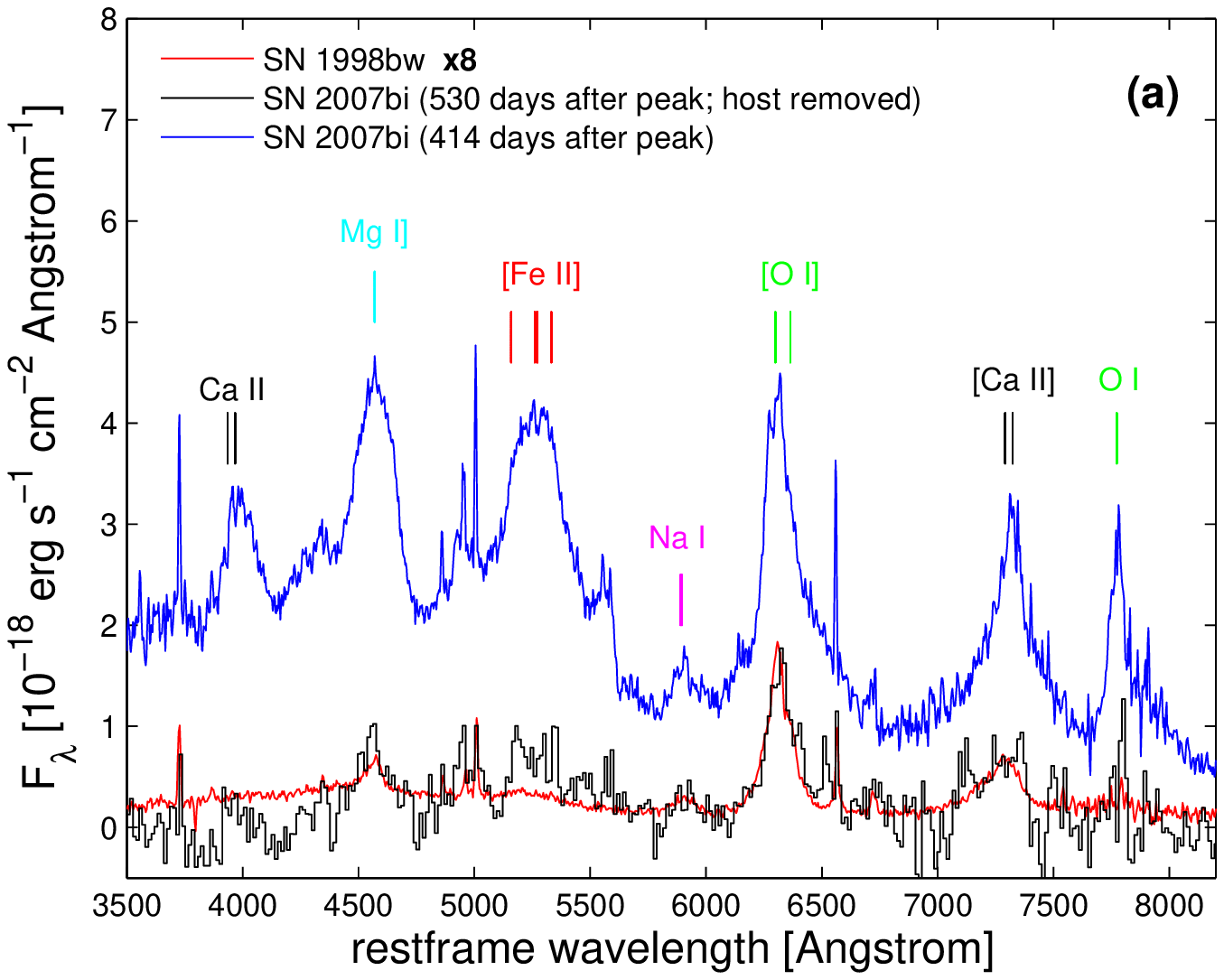}}
\centerline{\includegraphics[width=6in,angle=0]{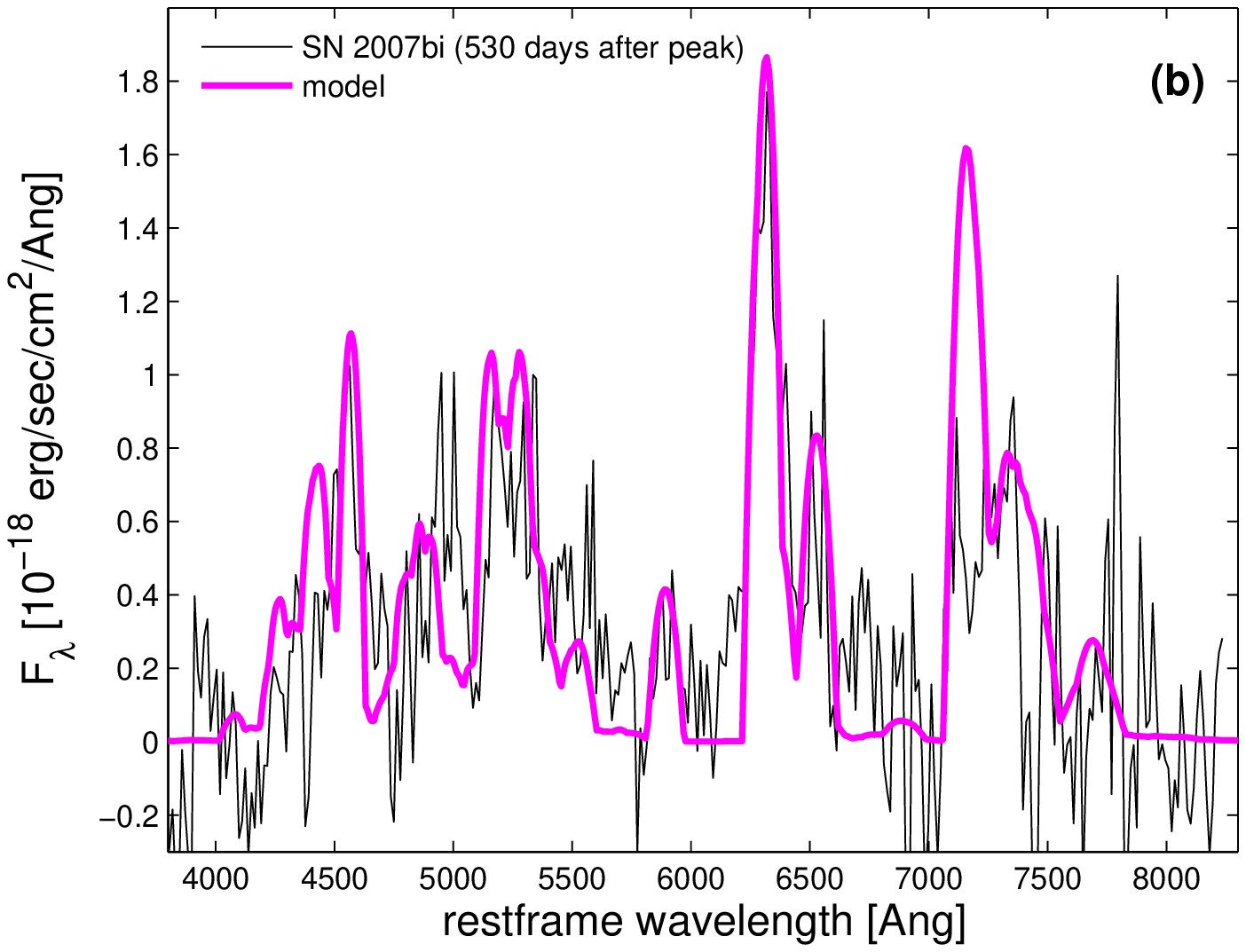}}
\bigskip
\caption[]{~\label{specevolfig}~}
\end{figure}

\clearpage

\begin{centering}
\underline{\Large Supplementary Information}\\
\end{centering}

~\\

\noindent{\bf (1) Technical observational details}\\

\noindent\underline{Photometry:}\\

Discovery and follow-up observations of SN 2007bi were obtained using the
Palomar-QUEST camera mounted on the 48'' Oschin Schmidt telescope at Palomar
Observatory (P48) as part of the SN Factory (SNF) program\cite{ald+2009}. 
These $R$-band observations were pipeline-reduced by the SNF
software, including bias removal, flatfield corrections and an astrometric
solution.   

Observations using the robotic 60-inch telescope at Palomar Observatory (P60) were 
pipeline-processed\cite{cen+2006}, including 
trimming, bias removal, flatfield corrections, and an astrometric solution. 

Observations using the 200-inch Hale telescope at Palomar Observatory (P200)
were obtained using the Large Format Camera (LFC) in SDSS $r$-band and were
cross-calibrated onto standard $R$ band as detailed below. 

Observations by the Catalina Sky Survey (CSS[15]) were obtained using
the CCD camera mounted on the 0.7\,m Catalina telescope. These unfiltered 
data were cross-calibrated onto the $R$-band grid as detailed below. 

Synthetic photometry was derived from the late-time Keck spectrum using the 
methods of ref. \cite{poz+2002}. 

Table 3 provides the full list of photometric data.\\ 

\noindent\underline{Spectroscopy:}\\

Early-time spectroscopy presented in Fig. 1\cite{nug+2007} was obtained with the
Low Resolution Imaging Spectrometer (LRIS[26])
mounted on the 10\,m Keck I telescope on Mauna Kea, Hawaii. The presented 
spectrum was obtained on Apr. 16, 2007 in long-slit mode. The exposure time
was 600\,s at airmass 1.08 under clear sky conditions with 
variable seeing around $2''$. The D560
dichroic was used with the 600 line mm$^{-1}$ grism on the blue side, and the
400 line mm$^{-1}$ grating blazed at 8500\,\AA\ on the red side, with the $1.5''$
slit oriented at the parallactic angle\cite{fil1982}. 
The spectral resolution achieved was $\sim 9$\,\AA\ on the red side and
$\sim 5$\,\AA\ on the blue side. Shutter problems cast doubt on the reliability
of the absolute flux calibration, so we have forced the spectral shape to match
that of a lower signal-to-noise ratio spectrum obtained on the previous night
(April 15, 2007) using the same instrument in spectropolarimetry mode 
and an otherwise identical setup, for which a reliable flux calibration was obtained. 

The spectrum of SN 1999as shown in Fig. 1 was obtained $\sim 3$ weeks
after discovery. Details about these data will be presented elsewhere, see initial
report in ref. \cite{den+2001}.

Late-time LRIS spectra presented in Fig. 3 were 
processed using a pipeline developed by
one of us (MS\cite{eli+2008}). Following standard
pre-processing (e.g., overscan subtraction), the data are divided by a
normalized flatfield removing pixel-to-pixel sensitivity variations
and correcting for the different gains of the CCD amplifiers.
Cosmic rays are identified and removed using LACOSMIC\cite{van2001}. 
We perform sky subtraction by subtracting a two-dimensional sky frame constructed from
sub-pixel sampling of the background spectrum and a knowledge of the
wavelength distortions as determined from two-dimensional comparison 
lamp frames. We also perform a fringe-frame correction on the
red side. The two-dimensional frames are transformed to a constant
dispersion using comparison lamp exposures. The spectral extraction
is performed by tracing the object position on the CCD and using a
variance-weighted extraction in a seeing-matched aperture. An error
spectrum from the statistics of the photon noise is also extracted.
The wavelength calibration of each extracted spectrum is then adjusted slightly
using the position of the night-sky lines to account for any drift in
the wavelength solution. Telluric correction and relative flux
calibration is performed using spectrophotometric standard stars. We
then finally combine the two sides into a single calibrated spectrum.
We match the flux across the dichroic by defining narrow box filters on
either side of the dichroic, and then use a weighted mean to combine
the spectra, in the process re-binning to a constant 2\,\AA\ per pixel
resolution. The result is a contiguous object spectrum, together with
an error spectrum representing the statistical uncertainties in the
flux in each binned pixel. 

The late-time spectrum from the Very Large Telescope (VLT) presented in 
Fig. 3 was taken on April 10, 2008 using the FORS2 spectrograph 
with the 300V+20 grism. Four 3600\,s  exposures were obtained with the
slit oriented along the parallactic angle. The spectra were
pre-reduced (bias and flatfield corrected), extracted, and wavelength-
and flux-calibrated using standard tasks within IRAF\footnote{IRAF (Image Reduction and Analysis Facility) is distributed
by the National Optical Astronomy Observatories, which are operated
by AURA, Inc., under cooperative agreement with the
National Science Foundation}. The wavelength and
flux calibration was computed using comparison lamps and the standard star
Feige~56 observed with the same instrumental configuration. Telluric 
features were removed using the standard-star spectrum (observed at similar
airmass). The combined spectrum revealed contamination from the host
galaxy that was subsequently removed using an extracted spectrum on
the edge of the galaxy where the SN contribution was negligible.

The absolute flux level of the late VLT spectrum was calibrated with 
photometry obtained during the same night[19]. The late-time
Keck spectrum does not have contemporaneous photometry. However, the
fact that the derived synthetic photometry is consistent with 
a smooth extrapolation from the previous photometry, and that the
nebular spectral analysis of both the photometry-calibrated VLT spectrum
and the later Keck spectrum give very similar results, suggests that
the absolute flux level of the Keck spectrum is reasonable.   

\clearpage

\noindent{\bf (2) SN 2007bi photometry}\\

We placed our $R$-band P60 photometry onto an absolute grid using 
SDSS photometry of multiple nearby sources, and solving for the
zero-point offset and color corrections for individual images using
a least-squares-based solver (E. O. Ofek, in preparation). A similar method
was employed for the P200 $r$-band images. Photometry of the P48 data was
obtained using the SNF pipeline and the derived $R$-band
magnitude agrees well with other data. CSS photometry was calibrated
onto an $R$-band grid by applying a constant zero-point offset, and 
overlapping data agree well with Palomar photometry (Fig. 1). 
Comparison with standard-star photometry from ref. [19]
shows an offset of $\sim0.1$ mag which we attribute
to residual differences due to telescope and filter transmissions.
Table 3 provides our full set of $R$-band data. Additional
photometry will be given in a forthcoming publication. 

In view of the lack of evidence for extinction of SN 2007bi in its dwarf host[19],
including non-detection of Na~D absorption lines in our spectra, we estimate that 
host extinction is negligible and correct only for Galactic extinction of
$A_R=0.07$\,mag\cite{sch+1998} taken from NED. We use a distance modulus appropriate
for $z=0.1279$ in the standard cosmology ($H_0 = 71$\,km\,s$^{-1}$\,Mpc$^{-1}$;
$\Omega_{\rm m}=0.27$; $\Omega_{\Lambda}=0.73$) of $m-M=38.86$\,mag. 

We subtract the host-galaxy contamination from our photometry in the
following manner. We calculate the $R$-band host-galaxy magnitude
$R_{\rm gal}=22.54$\,mag using available SDSS measurements ($r_{\rm gal}=22.62$\,mag,
$g_{\rm gal}=22.45$\,mag) from which we derive (using the formulae of ref. \cite{jor+2006})
$V-R=-0.0188$\,mag, $r-R=0.083$\,mag, and therefore $R_{\rm gal}=22.62-0.083=22.54$\,mag. 
We then subtract the host-galaxy flux from all photometric data.

Since the redshift of SN 2007bi is non-negligible, one needs in principle to
correct the observed $R$-band photometry into restframe $R$-band photometry
(so-called $K$-correction, $K_{RR}$). However, in view of the fact that
ref. [19] shows $K_{RR}<0.16$\,mag at all epochs, we neglect
this correction. 

To compare with models and other SNe it is useful to convert observed filtered
photometry into bolometric photometry. Combining our optical data, early-time
IR data (to be presented elsewhere) and data from ref. [19], we
find that the bolometric correction ($BC_R = M_{bol}-M_R$) is typically
$\sim -0.5$\,mag and is always $> -0.75$\,mag. When appropriate,
we use the range $0 > BC_R > -0.75$\,mag in our calculations.  

To estimate the SN peak magnitude and rise time we fit our sparse early-time
photometry with low-order polynomials. Experimenting with various polynomials 
and data ranges used in the fits, we derive typical values for the 
rise time (defined here as the time required for the SN to rise by 5\,mag
to peak) of $t_{\rm rise}=77$\,days and a peak magnitude $M_R=-21.3 \pm 0.15$\,mag. We show
typical fits in Fig. 4 (red and blue curves), as well as the fit that
yields our shortest rise time (45\,days). Of course, we cannot constrain more
complex light curve shapes given the sparsity of the data.
We adopt $t_{\rm rise}=77$\,days
as our fiducial rise-time value, and use the range $45 < t_{\rm rise} < 110$\,days
(restframe $40 < t_{\rm rise} < 97.5$\,days)
when appropriate (110\,days is a typical value derived from PISN models, Fig. 2b).

\clearpage 

%
\centerline{\includegraphics{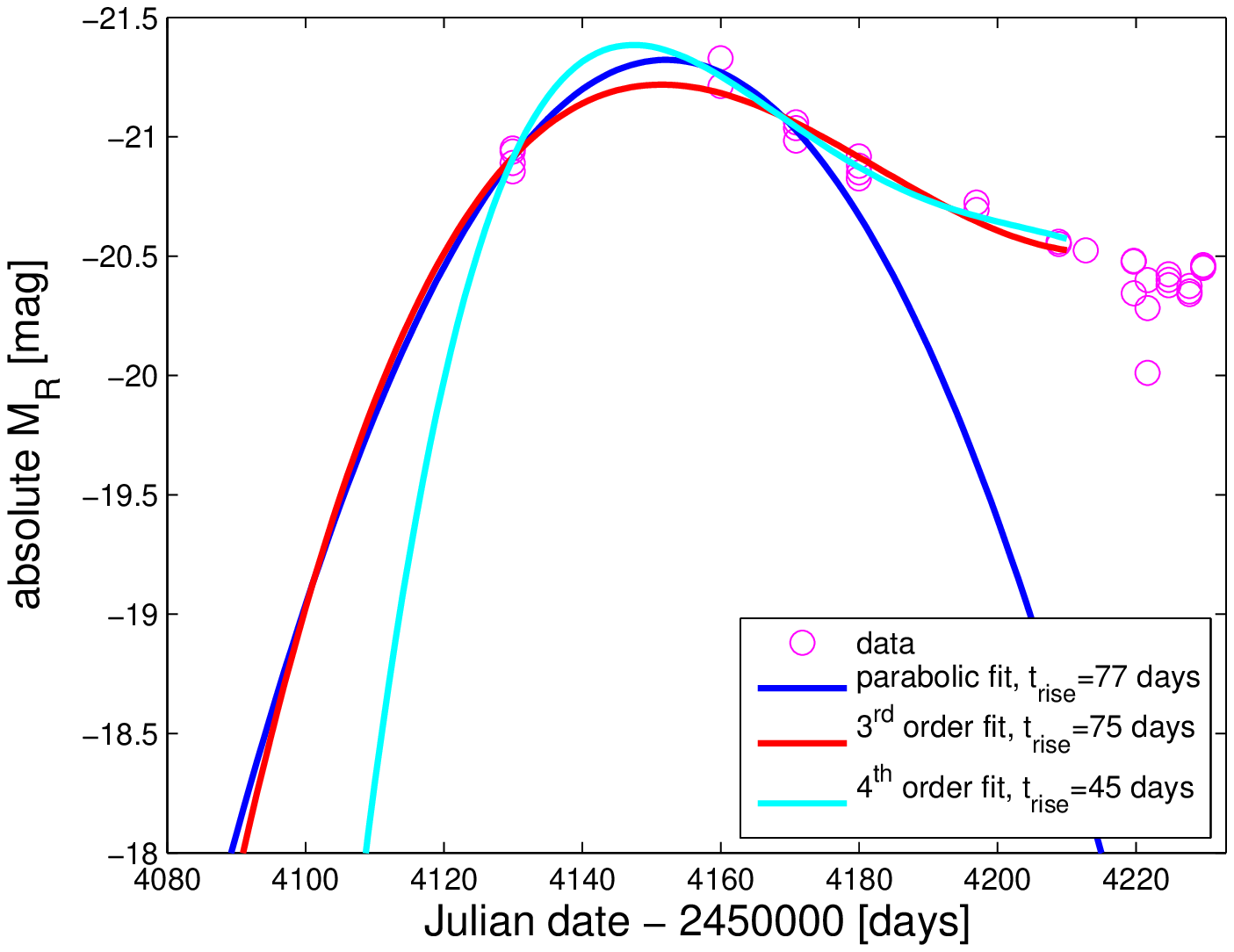}}
\bigskip

\noindent{\bf Figure 4}:\\

Polynomial fits to the early light-curve data. Polynomials
are fit to the data in a least-squares sense. Due to the paucity of the
data, higher order ($>5$) polynomials are underconstrained. 

\clearpage

\noindent{\bf (3) Derived physical properties}\\

We derive the physical properties of the explosion from the
observations using several independent methods. Our results 
are summarized in Table 2. \\

\noindent\underline{$^{56}$Ni mass:}\\

We estimate the synthesized $^{56}$Ni mass from the observed peak 
magnitude using the peak luminosity vs. $^{56}$Ni mass correlation[28]
and find $M_{^{56}{\rm Ni}}=3.5$\,M$_{\odot}$. 

An independent estimate for the $^{56}$Ni mass is obtained from the
luminosity of the radioactive tail of the light curve of SN 2007bi, which 
closely follows the theoretical decay slope ($0.0098$\,mag\,day$^{-1}$) 
of $^{56}$Co (the radioactive daughter product of $^{56}$Ni). We used a direct
comparison with the measured bolometric light curve of SN 1987A 
which is known to have been powered by the decay of $0.07$\,M$_{\odot}$
of $^{56}$Co at these stages[30] (Fig. 5).
We derive $M_{^{56}{\rm Ni}}=5.3$\,M$_{\odot}$, with a range of
$4.4$\,M$_{\odot} <M_{^{56}{\rm Ni}}<7$\,M$_{\odot}$ corresponding to 
restframe rise times $40$\,days $< t_{\rm rise} < 97.5$\,days and bolometric
corrections $-0.75$\,mag $< BC_{R} < 0$\,mag (see above).

Modelling of the nebular spectra of SN 2007bi (Fig. 3b, see also below)
yields estimates of the initial $^{56}$Ni mass since continued radioactivity
energizes the expanding ejecta and excites the observed emission lines. 
Models of the two available late-time spectra assuming a range of explosion dates
give estimates of  $3.7$\,M$_{\odot} <M_{^{56}{\rm Ni}}<7.4$\,M$_{\odot}$
(see main text, Table 1). 

Scaling our latest spectrum to nebular spectra of the well-studied SN 1998bw
(Fig. 2a) provides an estimate of the $^{56}$Ni mass assuming that the strength 
of the emission lines from the intermediate-mass elements (O, Mg, and Ca) is proportional
to $M_{^{56}{\rm Ni}}$. Correcting for the different restframe epochs of the spectra 
-- times since explosion of SN 1998bw (373 days) and SN 2007bi (470 days) -- we get
an estimated $^{56}$Ni mass range of $7$\,M$_{\odot} <M_{^{56}{\rm Ni}}<11$\,M$_{\odot}$
corresponding to restframe rise times in the range 
$40$\,days $< t_{\rm rise} < 97.5$\,days and 
assuming that SN 1998bw produced $0.5$\,M$_{\odot}$ of $^{56}$Ni[17]. Note
that this scaling is rough given the different O/Fe ratios of SNe 1998bw and 2007bi. 

Inspecting theoretical models[5,9] (Fig. 2b), we find that our
measured light curve is best fit by models producing 
$3$\,M$_{\odot} <M_{^{56}{\rm Ni}}<11$\,M$_{\odot}$ (total helium core 
masses $95$\,M$_{\odot} < M_{\rm He} < 110$\,M$_{\odot}$).\\

\noindent\underline{Total ejected mass:}\\

Modelling of the nebular spectra of SN 2007bi (Fig. 3b, see also below)
yields estimates of the total ejected mass of $51$\,M$_{\odot} <M_{\rm ej}<61$\,M$_{\odot}$
(see main text, Table 1). Note that these are strictly lower limits on the total 
mass for the following reasons. The velocity of the material emitting the
nebular emission lines ($v_{\rm neb}\approx5,600$\,km\,s$^{-1}$) is much lower than
the velocities observed during the very extended ($>100$\,days) photospheric 
phase ($v_{\rm ph}=12,000$\,km\,s$^{-1}$; Fig. 1, see also Fig. 16 of ref. [19].
This high-velocity material is therefore not contributing to the spectrum
at late times, and represents a substantial additional reservoir of mass in addition
to our derived values. In addition, almost all models of massive He cores predict that
the outer few solar masses of material will be composed mostly of He, while
we have not observed He lines at any epoch. This is usually explained by
the fact that He lines are non-thermally excited, and would only appear when
$^{56}$Ni is mixed into the He envelope providing a hard non-thermal exciting 
spectrum\cite{luc1991}. 
The lack of observed helium in SN 2007bi is therefore consistent with a spatial
segregation between an external He envelope and an internal core of heavier elements
into which newly synthesized $^{56}$Ni is mixed. The mass of this outer envelope
(which should be several solar masses) should again be added to the spectroscopic 
estimate given here. See below ($\S~6$) for additional discussion of the ejecta geometry. 

We can derive the total ejected mass $M_{\rm ej}$ in the explosion from commonly 
used[28,29] scaling relations. The required measurements
are the rise time $t_{\rm rise}$ and the photospheric velocity $v_{\rm ph}$. 
We note that these are
rough estimates that depend on the object used to anchor the scaling. 
Here, we compare SN 2007bi with two well-studied SNe~Ib/c for which the
rise time is known (owing to their coincidence with GRB/X-ray-flash events
that fix the explosion time): SN 1998bw and SN 2008D. 
We adopt $v_{\rm ph}=20,000$\,km\,s$^{-1}$, $t_{\rm rise}=17$\,days, and 
$M_{\rm ej}=11$\,M$_{\odot}$ for SN 1998bw[17]; 
$v_{\rm ph}=10,000$\,km\,s$^{-1}$, $t_{\rm rise}=19$\,days, 
and $M_{\rm ej}=7$\,M$_{\odot}$ for SN 2008D\cite{maz+2008}; 
and $v_{\rm ph}=12,000$\,km\,s$^{-1}$ (Fig. 1) and $t_{\rm rise}=66$\, 
restframe days (with a range $40$\,days $< t_{\rm rise} < 97.5$\,days)
for SN 2007bi (see above). Using the relations from ref. [29] we get
an estimated mass of $M_{\rm ej}\approx105$\,M$_{\odot}$ (with a range 
$36$\,M$_{\odot} <M_{ej}<173$\,M$_{\odot}$ for 
restframe rise times $40$\,days $< t_{\rm rise} < 97.5$\,days) and almost exactly the
same results using either SN 1998bw or SN 2008D. We note that the lower range
of these values is inconsistent with the spectroscopic estimates provided above,
arguing against such shorter rise times (i.e., 
indicating that  $t_{\rm rise} > 60$\,days).

Inspecting theoretical models[5,9] (Fig. 2b), we find that our
measured light curve is best fit by models with total helium core masses 
$95$\,M$_{\odot} < M_{\rm He} < 110$\,M$_{\odot}$.\\

\noindent\underline{Kinetic energy:}\\

We can derive the kinetic energy $E_{\rm k}$ generated by the explosion from 
commonly used[28,29] scaling relations. The required measurements
are the rise time $t_{\rm rise}$ and the photospheric velocity $v_{\rm ph}$. 
We note that these are
rough estimates that depend on the object used to anchor the scaling. 
Here, we compare SN 2007bi with two well-studied SNe~Ib/c for which the
rise time is known, as discussed above: SN 1998bw and SN 2008D. 
We adopt $v_{\rm ph}=20,000$\,km\,s$^{-1}$, $t_{\rm rise}=17$\,days, and 
$E_{\rm k}=30 \times 10^{51}$\,erg
for SN 1998bw[17]; $v_{\rm ph}=10,000$\,km\,s$^{-1}$, 
$t_{\rm rise}=19$\,days, and $E_{\rm k}=6\times 10^{51}$\,erg for 
SN 2008D\cite{maz+2008}; and $v_{\rm ph}=12,000$\,km\,s$^{-1}$ (Fig. 1) and 
$t_{\rm rise}=66$\, restframe days (with a range $40$\,days $< t_{\rm rise} < 97.5$\,days)
for SN 2007bi (see above). Using the relations from ref. [29] we get
an estimated energy of $E_{\rm k}\approx115\times 10^{51}$\,erg (with a range 
$36\times 10^{51}$\,erg $<E_{\rm k}<273\times 10^{51}$\,erg for 
restframe rise times $40$\,days $< t_{\rm rise} < 97.5$\,days) and again, similar
results using either SN 1998bw or SN 2008D. We note that the upper range
of these values appears to exceed the total budget of available nuclear 
energy in a PISN ($E<80\times 10^{51}$\,erg)[5], 
arguing for shorter rise times. It therefore 
appears that our fiducial rise time (observed $t_{\rm rise} \approx 77$\,days; 
$t_{\rm rise} = 66$\,days at restframe) provides
a reasonable fit considering all available constraints. 

As a sanity check we can directly estimate the kinetic energy using 
$E_{\rm k} = (1/2)M_{\rm ej}\, \bar{v}^2$, where $\bar{v}$ is the mass-averaged
expansion velocity. Assuming our fiducial estimated mass $M_{\rm ej}=100$\,M$_{\odot}$,
and that $\sim 1/2$ of that mass lies at low (nebular) velocities based
on our nebular analysis while the rest travels at velocities around
the photospheric values, we can adopt as the mean $\bar{v}=8,000$\,km\,s$^{-1}$
and derive $E_{\rm k}\approx80\times 10^{51}$\,erg.\\

\noindent\underline{Radiated energy:}\\

Direct integration under the observed light curve provides an estimate of the
total radiated energy of $1\times 10^{51}$\,erg $<E_{\rm rad}<2\times 10^{51}$\,erg
for a range of bolometric
corrections $-0.75$\,mag $< BC_{R} < 0$\,mag (see above). This is comparable 
to the total luminosity of the brightest SNe known[7] (see 
immediately below). 

\begin{table}
\begin{scriptsize}
\begin{tabular}{|c|c|c|c|}
\hline
Quantity & Method & Value [range] & Assumptions\tabularnewline
\hline
\hline
$^{56}$Ni mass & Peak magnitude & 3.5\,M$_{\odot}$ & \tabularnewline
               & SN 1987A comparison & 5.3 [4.4 .. 7]\,M$_{\odot}$ & $t_{rise}=[45 .. 110]$\,days, $BC_{R}=[-0.75 ..1]$\,mag \tabularnewline
               & Nebular modelling & [3.7 .. 7.4]\,M$_{\odot}$ &  $t_{rise}=[45 .. 110]$\,days\tabularnewline
               & SN 1998bw comparison & 8.9 [7.7 .. 11.3]\,M$_{\odot}$ & $t_{rise}=[45 .. 110]$\,days\tabularnewline
               & Light-curve models & [2.7 .. 11]\,M$_{\odot}$ & ref. [5]\tabularnewline
\hline
\hline
Ejected mass   & Nebular modelling & $>50$\,M$_{\odot}$ &  $t_{rise}=[45 .. 110]$\,days \tabularnewline
               & Light-curve scaling & 105 [37 .. 173]\,M$_{\odot}$ & $t_{\rm rise}=[45 .. 110]$\,days\tabularnewline
               & Light-curve models & [95 .. 110]\,M$_{\odot}$ & ref. [5]\tabularnewline
\hline
\hline
Kinetic energy & Light curve scaling & 132 [68 .. 273] $10^{51}$\,erg & $t_{\rm rise}=[45 .. 110]$\,days\tabularnewline
               & (1/2)$M_{\rm ej} \times \bar{v}^2$ & 80 $10^{51}$\,erg & $M_{\rm ej}=100$\,M$_{\odot}, \bar{v}=8,000$\,km\,s$^{-1}$ \tabularnewline
\hline
\hline
Radiated energy & Direct integration & [1 .. 2] $10^{51}$\,erg &  $BC_{R}=[-0.75 ..1]$\,mag \tabularnewline
\hline
\end{tabular}
\end{scriptsize}
\caption{Summary of physical properties derived}
\end{table}

\clearpage

\noindent{\bf (4) Comparison with other luminous SNe}\\

Several very luminous SNe were reported in the last
few years, and have been speculated to be PISNe.
Most prominent were SN 2006gy\cite{ofe+2007,smi+2007,smi+2008-06gy} (Fig. 5),
SN 2005ap\cite{qui+2007}, SN 2006tf\cite{smi+2008},
and SN 2008es[27],\cite{gez+2009}. 
Two major differences distinguish SN 2007bi from these previous
cases and strongly suggest it was a PISN.
First, in all previous cases the spectra of these luminous
SNe showed evidence for hydrogen, which is lacking
in SN 2007bi. This has a fundamental implication, since
in the case of SN 2007bi, all of the observed ejecta
had to come from the helium core, allowing us to directly
constrain its mass, and ultimately to provide compelling
evidence for a PISN (a helium core mass above $50$\,M$_{\odot}$).
In contrast, the ejecta mass in other SNe may be dominated
by hydrogen, complicating an attempt to constrain the
helium core mass.

Next, some of the previously studied luminous SNe showed
strong signatures of CSM 
interaction\cite{ofe+2007},\cite{smi+2008},[7],[14],[27], 
which SN 2007bi lacks.
Thus, the luminosity of SN 2007bi reflects directly
on the physics of the explosion (radioactive element
synthesis, explosion energy). The luminosity in other
cases may be dominated by conversion of kinetic energy
from a more standard SN explosion into luminosity via
shocks launched following a collision with a massive
CSM, previously lost from the progenitor star. Thus, while
the accumulated data for SN 2006gy seem to converge on
an extremely massive progenitor ($M = 100$\,M$_{\odot}$ or
more)[7,8], they disfavor a PISN
as the underlying energy source. In other cases (SN 2005ap\cite{qui+2007};
SN 2008es\cite{gez+2009},[27]), signatures of
interaction are less clear, but CSM interaction is
still favored as the source of the observed luminosity[27].

Compared to other hydrogen- and helium-deficient SNe of Type
Ic, SN 2007bi is by far the most luminous and energetic,
with the exception of SN 1999as[13],\cite{den+2001},\cite{hat+2001}
which appears quite similar, but lacks
observations (especially late-time photometry and spectroscopy)
of similar quality. Even the most energetic ``hypernovae''
associated with GRBs\cite{gal+1998,sta+2003,hjo+2003}
pale in comparison, with approximately an order of magnitude
less energy released and radioactive $^{56}$Ni produced[17] (Fig. 5).
Indeed, the term ``hypernovae'' seems better suited to
the truly extreme PISN explosions we have identified 
here\cite{wos+1981}.

\clearpage

%
\centerline{\includegraphics[width=5in]{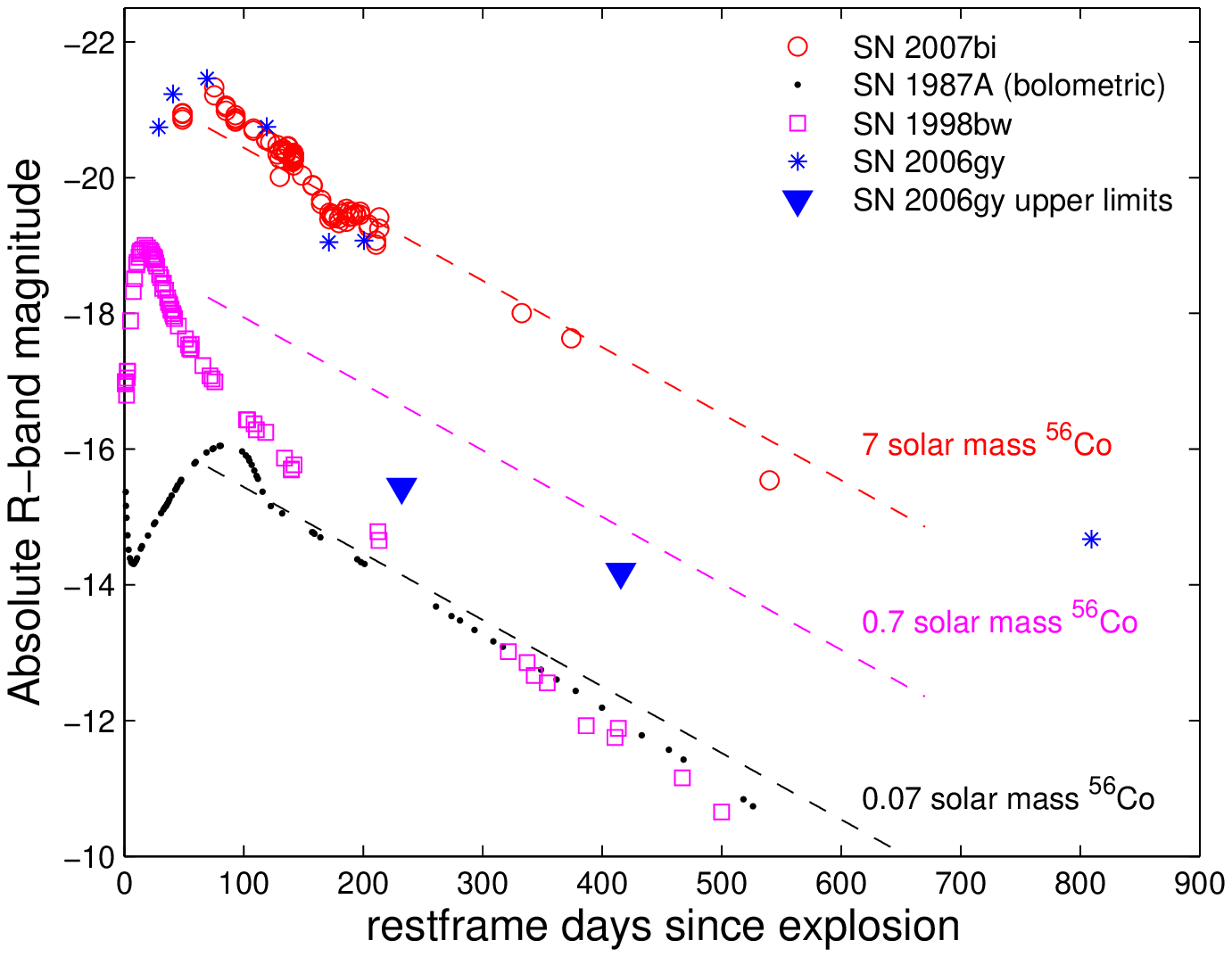}}
\bigskip

\noindent{\bf Figure 5}:

\begin{scriptsize}
Comparison with other luminous SNe.
We compare the $R$-band light curve of SN 2007bi (red circles)
with that of other luminous SNe observed at late times: the very luminous
SN 2006gy[8,14] (blue stars) and the prototypical
GRB/hypernova SN 1998bw\cite{gal+1998,pat+2001,mck+1999,sol+2000}
(magenta squares). We anchor our discussion on the well-studied SN 1987A
whose bolometric light curve[30] is also presented (black
dots). SN 1987A produced $\sim 0.07$\,M$_{\odot}$ of $^{56}$Ni.
The decay of its daughter nucleus $^{56}$Co drives the late-time
emission, as can be seen from the comparison with the theoretical
decay line plotted as the dashed black curve. While the relatively massive
and slow ejecta of SN 1987A generally provide an efficient envelope to
trap the radioactive energy and convert it into radiation, the
fact that the observations
slowly fall below the $^{56}$Co line indicates that, as time goes by,
more and more energy leaks out
of the expanding remnant without contributing to the radiative output.
Comparing the much more luminous SN 1998bw (magenta) to SN 1987A, one sees
that the much higher $^{56}$Ni production drives a more luminous
peak, but the lower ratio of kinetic energy to ejected mass from the stripped 
progenitor results in an inefficient trapping of the radioactive energy released, and
the observed light curves falls rapidly below the energy release rate.
Even more luminous SN 2006gy (magenta stars) would require $>7$\,M$_{\odot}$
of $^{56}$Ni to reach the observed peak. However, as can be seen from
comparison between the expected slow decay and the deep non-detections
at late times (magenta inverted triangles), the observations are
inconsistent with a radioactively driven evolution. Instead, this event
is now understood as resulting from strong CSM 
interaction[7],[8],[14],\cite{ofe+2007},\cite{smi+2007},\cite{smi+2008-06gy}, 
though of an unconventional sort with the interaction region initially opaque and 
invisible\cite{smith-mccray-2007},
and the late-time luminosity came from a reflected-light echo[8].
Finally, as we have reported here, the evolution of SN 2007bi is fully
consistent with a PISN of an extremely massive star,
producing several solar masses of $^{56}$Ni (and
enough ejecta to trap the radioactive decay energy), 
driving the light curve out to very late times, in perfect concordance
with theoretical $^{56}$Co decay.
\end{scriptsize}

\clearpage

\noindent{\bf (5) Spectroscopic modelling}\\

\noindent\underline{Photospheric spectral fitting:}\\

In Fig. 1 we present an automatically derived\cite{tho+2009} SYNOW\cite{bra+2002} fit 
to our data. The best automatic fit derived agrees with our best manually derived
attempts, and indicates a photospheric velocity $v_{\rm ph}=12,000$\,km\,s$^{-1}$. 
Prominent lines of iron, calcium, and magnesium are seen, and flux depression in
the blue side of the spectrum results from iron-group element (Fe, Co, Ni) line blends,
as typically found for Type I SNe. Lines of neutral oxygen and sodium, which 
are often prominent in early Type Ic SN spectra, appear remarkably weak. Independent 
analysis by ref. [19] arrives at very similar results. \\

\noindent\underline{Nebular spectral fitting:}\\

In Fig. 2 (bottom) we compare our Keck nebular spectrum with models derived using a
well-tested nebular modelling code[17], operated in a single-zone mode. 
The code calculates radioactive excitation and nebular emission cooling using extensive
line lists and constrains the amounts of radioactivity and the mass of the various 
elements. Since cooling effects at all wavelengths (extending outside of the optical window)
are considered, elements lacking strong optical nebular lines are also
constrained. Since the same code was used to study SN 1998bw[17],
the relative results (SN 2007bi vs. SN 1998bw) are quite robust, and indicate that 
SN 2007bi produced a factor of $\simgt 10$ more $^{56}$Ni than SN 1998bw.

\clearpage

\noindent{\bf (6) The spatial structure of the ejecta at late times}\\

In Fig. 6 we present a schematic illustration of the 
apparent geometrical distribution of the ejecta. The main
feature is that radioactive $^{56}$Ni (decaying into $^{56}$Co
at late times) appears to be centrally concentrated, and not
to have been mixed all the way out into the outermost layers
of the envelope. It thus illuminates mainly the more slowly expanding,
heavy-element-rich inner ejecta. The outermost, faster layers are
not illuminated at late times, explaining the slower velocity of
the material emitting the nebular spectra, and probably its 
composition, which appears to be depleted in C, O, and Mg relative
to Fe. 
The outermost helium layer would probably lie even farther outside,
well away from most of the $^{56}$Ni synthesized, and thus
does not contribute to either early (photospheric) or late
(nebular) spectra. This simple spherical scheme appears to 
explain all available data, including the following:

{\bf (a)} The high mass
estimated from light-curve modelling, which is sensitive to
all of the material contributing to the opacity at the photospheric
phase, including the outer, faster shells, compared to the lower
total mass derived from nebular spectroscopic modelling, sensitive
only to slower, inner shells which are highly enriched in radioactive
material. 
 
{\bf (b)} The composition derived from the nebular spectrum which,
relative to iron,
is depleted in lighter elements (C, O, Mg) that are more abundant
in the outer layers of the envelope. 

{\bf (c)} The lack of helium lines at all times, which are
segregated from the energizing $^{56}$Ni and are thus not excited\cite{luc1991}. 

We note that the analysis of ref. [19] does not show evidence
for asphericity in late-time spectra of SN 2007bi as seen in many other
Type Ib/c events\cite{maz+2005,mae+2008,mod+2008,tau+2009}. Along with
the apparent suggestion that $^{56}$Ni is not well mixed, the data
probably argue against a bipolar/jet-driven explosion model as proposed for
normal and GRB-related SNe Ib/c.

\clearpage

%
\centerline{\includegraphics[width=5.5in,angle=-90]{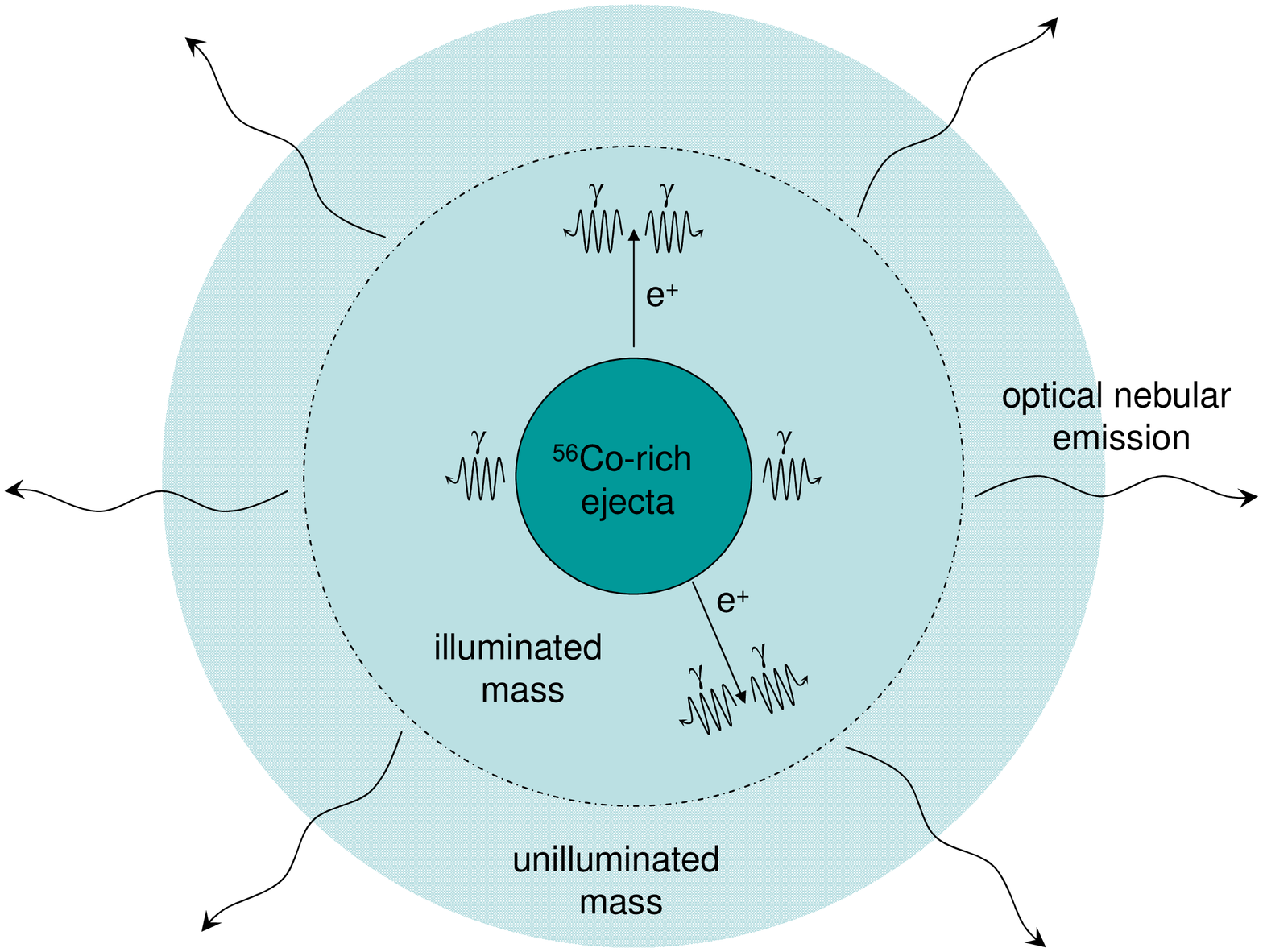}}
\bigskip

\noindent{\bf Figure 6}:

A schematic illustration of the ejecta geometry. $^{56}$Ni (decaying
to $^{56}$Co) is most abundant in the core, and emits positrons and gamma rays
that excite the surrounding material. The outer layers of the envelope, which
are expected to be dominated by lighter elements (C, O, Mg) and where any
helium must reside, are not well mixed with the radioactive elements and
thus remain unilluminated by hard radiation at late times, and do not contribute
to the nebular spectrum (nor to the derived mass and composition from the
analysis of these data).

\clearpage



\begin{table}
\begin{scriptsize}
\begin{tabular}{|c|c|c|c|c|c|}
\hline
Jd [day] & $R$ [mag] & Error [mag] & Jd [day] & $R$ [mag] & Error [mag]\tabularnewline
\hline
\hline
2454219.67 & 18.38 & 0.08 & 2454219.67 & 18.38 & 0.08\tabularnewline
2454219.67 & 18.52 & 0.08 & 2454221.68 & 18.46 & 0.08\tabularnewline
2454221.69 & 18.58 & 0.08 & 2454221.69 & 18.85 & 0.08\tabularnewline
2454224.73 & 18.44 & 0.08 & 2454224.73 & 18.48 & 0.08\tabularnewline
2454224.73 & 18.46 & 0.08 & 2454227.72 & 18.52 & 0.08\tabularnewline
2454227.72 & 18.48 & 0.08 & 2454227.72 & 18.51 & 0.08\tabularnewline
2454229.70 & 18.40 & 0.08 & 2454229.70 & 18.41 & 0.08\tabularnewline
2454229.70 & 18.40 & 0.08 & 2454234.66 & 18.60 & 0.08\tabularnewline
2454234.66 & 18.53 & 0.08 & 2454234.69 & 18.55 & 0.08\tabularnewline
2454234.70 & 18.60 & 0.08 & 2454234.70 & 18.59 & 0.08\tabularnewline
2454234.70 & 18.51 & 0.08 & 2454234.70 & 18.56 & 0.08\tabularnewline
2454234.71 & 18.55 & 0.08 & 2454234.71 & 18.54 & 0.08\tabularnewline
2454234.71 & 18.58 & 0.08 & 2454234.71 & 18.60 & 0.08\tabularnewline
2454234.71 & 18.55 & 0.08 & 2454234.71 & 18.56 & 0.08\tabularnewline
2454234.72 & 18.58 & 0.08 & 2454234.72 & 18.59 & 0.08\tabularnewline
2454234.72 & 18.55 & 0.08 & 2454234.72 & 18.59 & 0.08\tabularnewline
2454234.72 & 18.52 & 0.08 & 2454234.72 & 18.54 & 0.08\tabularnewline
2454234.72 & 18.59 & 0.08 & 2454234.73 & 18.57 & 0.08\tabularnewline
2454234.73 & 18.59 & 0.08 & 2454234.73 & 18.49 & 0.08\tabularnewline
2454234.73 & 18.52 & 0.08 & 2454234.73 & 18.52 & 0.08\tabularnewline
2454234.73 & 18.55 & 0.08 & 2454234.74 & 18.59 & 0.08\tabularnewline
2454234.74 & 18.61 & 0.08 & 2454234.74 & 18.58 & 0.08\tabularnewline
2454234.74 & 18.68 & 0.08 & 2454234.74 & 18.56 & 0.08\tabularnewline
2454234.74 & 18.59 & 0.08 & 2454252.67 & 18.97 & 0.08\tabularnewline
2454252.67 & 18.98 & 0.08 & 2454260.69 & 19.25 & 0.08\tabularnewline
2454260.69 & 19.19 & 0.08 & 2454268.71 & 19.38 & 0.08\tabularnewline
2454268.71 & 19.47 & 0.08 & 2454270.70 & 19.43 & 0.08\tabularnewline
2454270.70 & 19.40 & 0.08 & 2454272.68 & 19.45 & 0.08\tabularnewline

\hline
\end{tabular}
\end{scriptsize}
\caption{$R$-band photometry of SN 2007bi (Cont. next page)}
\end{table}

\clearpage

\begin{table}
\begin{scriptsize}
\begin{tabular}{|c|c|c|c|c|c|}
\hline
Jd [day] & $R$ [mag] & Error [mag] & Jd [day] & $R$ [mag] & Error [mag]\tabularnewline
\hline
\hline
2454272.69 & 19.41 & 0.08 & 2454277.68 & 19.53 & 0.08\tabularnewline
2454277.68 & 19.46 & 0.08 & 2454280.68 & 19.39 & 0.08\tabularnewline
2454284.76 & 19.32 & 0.08 & 2454284.76 & 19.51 & 0.08\tabularnewline
2454287.71 & 19.36 & 0.08 & 2454287.71 & 19.44 & 0.08\tabularnewline
2454290.69 & 19.43 & 0.08 & 2454290.69 & 19.37 & 0.08\tabularnewline
2454293.71 & 19.41 & 0.08 & 2454293.71 & 19.39 & 0.08\tabularnewline
2454297.70 & 19.41 & 0.08 & 2454297.70 & 19.36 & 0.08\tabularnewline
2454305.67 & 19.55 & 0.08 & 2454305.67 & 19.60 & 0.08\tabularnewline
2454312.69 & 19.85 & 0.08 & 2454312.69 & 19.79 & 0.08\tabularnewline
2454315.66 & 19.45 & 0.08 & 2454315.66 & 19.62 & 0.08\tabularnewline
2454497.00 & 21.23 & 0.50 & 2454684.00 & 23.32 & 0.60\tabularnewline
2454196.95 & 18.17 & 0.11 & 2454196.97 & 18.14 & 0.13\tabularnewline
2454208.81 & 18.30 & 0.05 & 2454208.85 & 18.31 & 0.03\tabularnewline
2454129.93 & 17.92 & 0.06 & 2454129.93 & 18.00 & 0.06\tabularnewline
2454129.94 & 17.91 & 0.06 & 2454129.95 & 17.97 & 0.06\tabularnewline
2454159.95 & 17.65 & 0.06 & 2454159.97 & 17.53 & 0.07\tabularnewline
2454170.87 & 17.82 & 0.06 & 2454170.87 & 17.80 & 0.06\tabularnewline
2454170.88 & 17.88 & 0.06 & 2454179.96 & 17.94 & 0.06\tabularnewline
2454179.97 & 18.03 & 0.07 & 2454179.97 & 17.98 & 0.07\tabularnewline
2454179.98 & 18.01 & 0.07 & 2454212.74 & 18.34 & 0.12\tabularnewline
2454233.77 & 18.63 & 0.07 & 2454242.70 & 18.83 & 0.20\tabularnewline
2454450.02 & 20.86 & 0.47 & & & \tabularnewline
\hline
\end{tabular}
\end{scriptsize}
\end{table}



\begin{thebibliography}{100}

\bibitem[1]{sma2009}
{Smartt}, S.~J.  {Progenitors of Core-Collapse Supernovae}.
\newblock {Preprint at: ${\rm <http://arxiv.org/abs/0908.0700>}$} (2009).

\bibitem[2]{gal+2009}
{{Gal-Yam}, A. \& {Leonard}, D.~C.} {A massive hypergiant star as the progenitor of the supernova SN 2005gl}.
\newblock {\it Nature} {\bf 458}, 865--867 (2009).

\bibitem[3]{rak+1967}
{{Rakavy}, G. \& {Shaviv}, G.} {Instabilities in Highly Evolved Stellar Models}.
\newblock {\it Astrophys. J.} {\bf 148}, 803 (1967).

\bibitem[4]{bar+1967}
{{Barkat}, Z., {Rakavy}, G. \& {Sack}, N.} {Dynamics of Supernova Explosion Resulting from Pair Formation}.
\newblock {\it Phys. Rev. Lett.} {\bf 18}, 379--381 (1967).

\bibitem[5]{heg+2002}
{{Heger}, A. \& {Woosley}, S.~E.} {The Nucleosynthetic Signature of Population III}.
\newblock {\it Astrophys. J.} {\bf 567}, 532--543 (2002).

\bibitem[6]{wos+2007}
{{Woosley}, S.~E., {Blinnikov}, S. \& {Heger}, A.} {Pulsational pair instability as an explanation for the most luminous supernovae}.
\newblock {\it Nature} {\bf 450}, 390--392 (2007).

\bibitem[7]{smi+2009}
{{Smith}, N., {Chornock}, R., {Silverman}, J.~M., {Filippenko}, A.~V. \&
 {Foley}, R.~J.} {Spectral Evolution of the Extraordinary Type IIn Supernova 2006gy}.
\newblock {Preprint at: ${\rm <http://arxiv.org/abs/0906.2200>}$} (2009).

\bibitem[8]{mil+2009b}
{{Miller}, A.~A., {Smith}, N., {Li}, W., {Bloom}, J.~S.,
{Chornock}, R. {\it et al.}}
{New Observations of the Very Luminous Supernova 2006gy: Evidence for Echoes}.
\newblock {Preprint at: ${\rm <http://arxiv.org/abs/0906.2201>}$} (2009).

\bibitem[9]{kas+2008}
{{Kasen}, D., {Heger}, A. \& {Woosley}, S.}
{The First Stellar Explosions: Theoretical Light Curves and Spectra of Pair-Instability Supernovae}.
\newblock {American Institute of Physics Conference Series} {\bf 990}, 263--267 (2008).

\bibitem[10]{wal2008}
{{Waldman}, R.} {The Most Massive Core-Collapse Supernova Progenitors}.
\newblock {\it Astrophys. J.} {\bf 685}, 1103--1108 (2008).

\bibitem[11]{fig2005}
{{Figer}, D.~F.} {An upper limit to the masses of stars}.
\newblock {\it Nature} {\bf 434}, 192--194 (2005).

\bibitem[12]{fil97}
{Filippenko}, A.~V. {Optical spectra of supernovae}.
\newblock {\it Annu. Rev. Astron. Astr.} {\bf 35}, 309--355 (1997).

\bibitem[13]{kas2004}
{Kasen}, D.~N. {Aspherical supernovae}.
\newblock {PhD Thesis -- University of California, Berkeley, California, USA} (2004).

\bibitem[14]{agn+2009}
{{Agnoletto}, I., {Benetti}, S., {Cappellaro}, E., {Zampieri}, L.,
 {Turatto}, M. {\it et al.}} {SN 2006gy: Was it Really Extraordinary?}
\newblock {\it Astrophys. J.} {\bf 691}, 1348--1359 (2009).

\bibitem[15]{dra+2009}
{{Drake}, A.~J., {Djorgovski}, S.~G., {Mahabal}, A.,
 {Beshore}, E., {Larson}, S. {\it et al.}} {First Results from the Catalina Real-Time Transient Survey}.
\newblock {\it Astrophys. J.} {\bf 696}, 870--884 (2009).

\bibitem[16]{cen+2009}
{{Cenko}, S.~B., {Frail}, D.~A., {Harrison}, F.~A.,
 {Kulkarni}, S.~R., {Nakar}, E. {\it et al.}}
{The Collimation and Energetics of the Brightest Swift Gamma-Ray Bursts}.
\newblock {Preprint at: ${\rm <http://arxiv.org/abs/0905.0690>}$} (2009).

\bibitem[17]{maz+2001}
{{Mazzali}, P.~A., {Nomoto}, K., {Patat}, F. \& {Maeda}, K.} {The Nebular Spectra of the Hypernova SN 1998bw and Evidence for Asymmetry}.
\newblock {\it Astrophys. J.} {\bf 559}, 1047--1053 (2001).

\bibitem[18]{lan+2007}
{{Langer}, N., {Norman}, C.~A., {de Koter}, A., {Vink}, J.~S., 
 {Cantiello}, M. {\it et al.}} {Pair creation supernovae at low and high redshift}.
\newblock {\it Astron. Astrophys.} {\bf 475}, L19--L23 (2007).

\bibitem[19]{you+2009}
{{Young}, D.~R., {Smartt}, S.~J., {Valenti}, S., {Pastorello}, A., {Benetti}, S. 
{\it et al.}} {Two peculiar type Ic supernovae in low-metallicity, dwarf galaxies}.
\newblock {\it Astron. Astrophys.} submitted (2009).

\bibitem[20]{ume+2008}
{{Umeda}, H. \& {Nomoto}, K.} {How Much $^{56}$Ni Can Be Produced in Core-Collapse Supernovae? Evolution and Explosions of 30-100 M$_{\odot}$ Stars}.
\newblock {\it Astrophys. J.} {\bf 673}, 1014--1022 (2008).

\bibitem[21]{bro+2004}
{Bromm}, V. \& {Larson}, R.~B. {The First Stars}.
\newblock {\it Annu. Rev. Astron. Astr.} {\bf 42}, 79--118 (2004).

\bibitem[22]{tre+2004}
{{Tremonti}, C.~A., {Heckman}, T.~M., {Kauffmann}, G.,
 {Brinchmann}, J., {Charlot}, S. {\it et al.}} {The Origin of the Mass-Metallicity Relation: Insights from 53,000 Star-forming Galaxies in the Sloan Digital Sky Survey}.
\newblock {\it Astrophys. J.} {\bf 613}, 898--913 (2004).

\bibitem[23]{law+2009}
{{Law}, N.~M., {Kulkarni}, S.~R., {Dekany}, R.~G., {Ofek}, E.~O.,
 {Quimby} {\it et al.}} {The Palomar Transient Factory: System Overview, Performance and First Results}.
\newblock {Preprint at: ${\rm <http://arxiv.org/abs/0906.5350>}$} (2009).

\bibitem[24]{rau+2009}
{{Rau}, A., {Kulkarni}, S.~R., {Law}, N.~M., {Bloom}, J.~S.,
 {Ciardi}, D., {Djorgovski}, G.~S. {\it et al.}} {Exploring the Optical Transient Sky with the Palomar Transient Factory}.
\newblock {Preprint at: ${\rm <http://arxiv.org/abs/0906.5355>}$} (2009).

\bibitem[25]{you+2008}
{{Young}, D.~R., {Smartt}, S.~J., {Mattila}, S., {Tanvir}, N.~R.,
 {Bersier}, D. {\it et al.}} {Core-collapse supernovae in low-metallicity environments and future all-sky transient surveys}.
\newblock {\it Astron. Astrophys.} {\bf 489}, 359--375 (2008).

\bibitem[26]{oke+1995}
{{Oke}, J.~B., {Cohen}, J.~G., {Carr}, M., {Cromer}, J., 
 {Dingizian}, A. {\it et al.}} {The Keck Low-Resolution Imaging Spectrometer}.
\newblock {\it Publ. Astron. Soc. Pac.} {\bf 107}, 375--385 (1995).

\bibitem[27]{mil+2009a}
{{Miller}, A.~A., {Chornock}, R., {Perley}, D.~A., {Ganeshalingam}, M.,
 {Li}, W. {\it et al.}} {The Exceptionally Luminous Type II-Linear Supernova 2008es}.
\newblock {\it Astrophys. J.} {\bf 690}, 1303--1312 (2009).

\bibitem[28]{per+2009}
{{Perets}, H.~B., {Gal-Yam}, A., {Mazzali}, P., {Arnett}, D.,
 {Kagan}, D. {\it et al.}} {A new type of stellar explosion}.
\newblock {Preprint at: ${\rm <http://arxiv.org/abs/0906.2003>}$} (2009).

\bibitem[29]{fol+2009}
{{Foley}, R.~J., {Chornock}, R., {Filippenko}, A.~V.,
 {Ganeshalingam}, M., {Kirshner}, R.~P. {\it et al.}} {SN 2008ha: An Extremely Low Luminosity and Exceptionally Low Energy Supernova}.
\newblock {\it Astron. J.} {\bf 138}, 376--391 (2009).

\bibitem[30]{pun+1995}
{{Pun}, C.~S.~J., {Kirshner}, R.~P., {Sonneborn}, G.,
 {Challis}, P., {Nassiopoulos}, G. {\it et al.}} {Ultraviolet Observations of SN 1987A with the IUE Satellite}.
\newblock {\it Astrophys. J. Suppl. Ser.} {\bf 99}, 223--261 (1995).

\end{thebibliography}

\begin{thebibliography}{100}

\bibitem[31]{ald+2009}
{{Aldering}, G.~S., {Antilogus}, P., {Aragon}, C., {Bailey}, S.,
 {Baltay} {\it et al.} }{SNe Ia From The Nearby Supernova Factory: The Reign Of The Normals And The Revolution Of The Rare}.
\newblock {{\it Bull. Am. Astron. Soc.} {\bf 41}, 401} (2009).

\bibitem[32]{cen+2006}
{{Cenko}, S.~B., {Fox}, D.~B., {Moon}, D.-S., {Harrison}, F.~A., 
{Kulkarni} {\it et al.}} {The Automated Palomar 60 Inch Telescope}.
\newblock {{\it Publ. Astron. Soc. Pac.} {\bf 118}, 1396--1406} (2006).

\bibitem[33]{poz+2002}
{{Poznanski}, D., {Gal-Yam}, A., {Maoz}, D., {Filippenko}, A.~V., 
 {Leonard}, D.~C. {\it et al.}} {Not Color-Blind: Using Multiband Photometry to Classify Supernovae}.
\newblock {{\it Publ. Astron. Soc. Pac.} {\bf 114}, 833--845} (2002).

\bibitem[34]{nug+2007}
{{Nugent}, P.~E.} {Supernova 2007bi}.
\newblock {{\it Central Bureau Electronic Telegrams} {\bf 929}} (2007).

\bibitem[35]{fil1982}
{{Filippenko}, A.~V.} {The importance of atmospheric differential refraction in spectrophotometry}.
\newblock {{\it Publ. Astron. Soc. Pac.} {\bf 94}, 715--721} (1982).

\bibitem[36]{den+2001}
{{Deng}, J., {Hatano}, K., {Nakamura}, T., {Maeda}, K., {Nomoto} {\it et al.}} {Luminous SN 1999as: Hypernova with 4Msun 56Ni Ejected?}.
\newblock {{\it ASP Conf. Proc. V.} {\bf 251}, 238} (2001).

\bibitem[37]{eli+2008}
{{Ellis}, R.~S., {Sullivan}, M., {Nugent}, P.~E., {Howell}, D.~A., 
{Gal-Yam}, A. {\it et al.}} {Verifying the Cosmological Utility of Type Ia Supernovae: Implications of a Dispersion in the Ultraviolet Spectra}.
\newblock {{\it Astrophys. J.} {\bf 674}, 51--69} (2008).

\bibitem[38]{van2001}
{{van Dokkum}, P.~G.} {Cosmic-Ray Rejection by Laplacian Edge Detection}.
\newblock {{\it Publ. Astron. Soc. Pac.} {\bf 113}, 1420--1427} (2001).

\bibitem[39]{sch+1998}
{{Schlegel}, D.~J., {Finkbeiner}, D.~P. \& {Davis}, M.} {Maps of Dust Infrared Emission for Use in Estimation of Reddening and Cosmic Microwave Background Radiation Foregrounds}.
\newblock {{\it Astrophys. J.} {\bf 500}, 525--553} (1998).

\bibitem[40]{jor+2006}
{{Jordi}, K., {Grebel}, E.~K. \& {Ammon}, K.} {Empirical color transformations between SDSS photometry and other photometric systems}.
\newblock {{\it Astron. Astrophys.} {\bf 460}, 339--347} (2006).

\bibitem[41]{luc1991}
{{Lucy}, L.~B.} {Nonthermal excitation of helium in type Ib supernovae}.
\newblock {{\it Astrophys. J.} {\bf 383}, 308--313} (1991).

\bibitem[42]{maz+2008}
{{Mazzali}, P.~A. and {Valenti}, S. and {Della Valle}, M. and
 {Chincarini}, G. and {Sauer}, D.~N. {\it et al}} {The Metamorphosis of Supernova SN 2008D/XRF 080109: A Link Between Supernovae and GRBs/Hypernovae}.
\newblock {{\it Science} {\bf 321}, 1185--1188} (2008).

\bibitem[43]{ofe+2007}
{{Ofek}, E.~O., {Cameron}, P.~B., {Kasliwal}, M.~M.,
 {Gal-Yam}, A., {Rau}, A. {\it et al}} {SN 2006gy: An Extremely Luminous Supernova in the Galaxy NGC 1260}.
\newblock {{\it Astrophys. J. Lett.} {\bf 659}, L13--L16} (2007).

\bibitem[44]{smi+2007}
{{Smith}, N., {Li}, W., {Foley}, R.~J., {Wheeler}, J.~C.,
 {Pooley}, D. {\it et al}} {SN 2006gy: Discovery of the Most Luminous Supernova Ever Recorded, Powered by the Death of an Extremely Massive Star like {$\eta$} Carinae}.
\newblock {{\it Astrophys. J.} {\bf 666}, 1116--1128} (2007).

\bibitem[45]{smi+2008-06gy}
{{Smith}, N., {Foley}, R.~J., {Bloom}, J.~S., {Li}, W., {Filippenko}, A.~V. {\it et al}} {Late-Time Observations of SN 2006gy: Still Going Strong}.
\newblock {{\it Astrophys. J.} {\bf 686}, 485--491} (2008).

\bibitem[46]{qui+2007}
{{Quimby}, R.~M., {Aldering}, G., {Wheeler}, J.~C., {H{\"o}flich}, P., {Akerlof}, C.~W. {\it et al}} {SN 2005ap: A Most Brilliant Explosion}.
\newblock {{\it Astrophys. J. Lett.} {\bf 668}, L99--L102} (2007).

\bibitem[47]{smi+2008}
{{Smith}, N., {Chornock}, R., {Li}, W., {Ganeshalingam}, M., {Silverman}, J.~M. {\it et al}} {SN 2006tf: Precursor Eruptions and the Optically Thick Regime of Extremely Luminous Type IIn Supernovae}.
\newblock {{\it Astrophys. J.} {\bf 686}, 467--484} (2008).

\bibitem[48]{gez+2009}
{{Gezari}, S., {Halpern}, J.~P., {Grupe}, D., {Yuan}, F., {Quimby}, R. {\it et al}} {Discovery of the Ultra-Bright Type II-L Supernova 2008es}.
\newblock {{\it Astrophys. J.} {\bf 690}, 1313--1321} (2009).

\bibitem[49]{hat+2001}
{{Hatano}, K., {Branch}, D., {Nomoto}, K., {Deng}, J.~S., {Maeda}, K. {\it et al.}} {The Type Ic Hypernova SN 1999as}.
\newblock {{\it Bull. Am. Astron. Soc.} {\bf 33}, 838} (2001).

\bibitem[50]{gal+1998}
{{Galama}, T.~J., {Vreeswijk}, P.~M., {van Paradijs}, J., {Kouveliotou}, C., {Augusteijn}, T. {\it et al.}} {An unusual supernova in the error box of the {$\gamma$}-ray burst of 25 April 1998}.
\newblock {{\it Nature} {\bf 395}, 670--672} (1998).

\bibitem[51]{sta+2003}
{{Stanek}, K.~Z., {Matheson}, T., {Garnavich}, P.~M., {Martini}, P., {Berlind}, P. {\it et al.}} {Spectroscopic Discovery of the Supernova 2003dh Associated with GRB 030329}.
\newblock {{\it Astrophys. J. Lett.} {\bf 591}, L17--L20} (2003).

\bibitem[52]{hjo+2003}
{{Hjorth}, J., {Sollerman}, J., {M{\o}ller}, P., {Fynbo}, J.~P.~U., {Woosley}, S.~E. {\it et al.}} {A very energetic supernova associated with the {$\gamma$}-ray burst of 29 March 2003}.
\newblock {{\it Nature} {\bf 423}, 847--850} (2003).

\bibitem[53]{wos+1981}
{Woosley, S. E. \& Weaver, T. A. {\it et al.}} {Theoretical models for supernovae}.
\newblock {{\it NATO Adv. Study Inst. on Supernovae}} (2001).

\bibitem[54]{pat+2001}
{{Patat}, F., {Cappellaro}, E., {Danziger}, J., {Mazzali}, P.~A., {Sollerman}, J. {\it et al.}} {The Metamorphosis of SN 1998bw}.
\newblock {{\it Astrophys. J.} {\bf 555}, 900--917} (2001).

\bibitem[55]{mck+1999}
{{McKenzie}, E.~H. \& {Schaefer}, B.~E.} {The Late-Time Light Curve of SN 1998bw Associated with GRB 980425}.
\newblock {{\it Publ. Astron. Soc. Pac.} {\bf 111}, 964--968} (1999).

\bibitem[56]{sol+2000}
{{Sollerman}, J., {Kozma}, C., {Fransson}, C., {Leibundgut}, B., {Lundqvist}, P. {\it et al.}} {SN 1998bw at Late Phases}.
\newblock {{\it Astrophys. J. Lett.} {\bf 537}, L127--L130} (2000).

\bibitem[57]{smith-mccray-2007}
{{Smith}, N. \& {McCray}, R.} {Shell-shocked Diffusion Model for the Light Curve of SN 2006gy}.
\newblock {{\it Astrophys. J. Lett.} {\bf 671}, L17--L20} (2007).

\bibitem[58]{tho+2009}
{{Thomas}, R., {Aldering}, G., {Antilogus}, P., {Aragon}, C.,
 {Bailey}, S. {\it et al.}} {Direct Analysis of Type Ia Supernovae Observed by The Nearby Supernova Factory}.
\newblock {{\it Bull. Am. Astron. Soc.} {\bf 41}, 464} (2009).

\bibitem[59]{bra+2002}
{{Branch}, D., {Benetti}, S., {Kasen}, D., {Baron}, E.,
 {Jeffery}, D.~J. {\it et al.}} {Direct Analysis of Spectra of Type Ib Supernovae}.
\newblock {{\it Astrophys. J.} {\bf 566}, 1005--1017} (2002).

\bibitem[60]{maz+2005}
{{Mazzali}, P.~A., {Kawabata}, K.~S., {Maeda}, K., {Nomoto}, K.,  {Filippenko}, A.~V. {\it et al.}} {An Asymmetric Energetic Type Ic Supernova Viewed Off-Axis, and a Link to Gamma Ray Bursts}.
\newblock {{\it Science} {\bf 308}, 1284--1287} (2005).

\bibitem[61]{mae+2008}
{{Maeda}, K., {Kawabata}, K., {Mazzali}, P.~A., {Tanaka}, M., {Valenti}, S.{\it et al.}} {Asphericity in Supernova Explosions from Late-Time Spectroscopy}.
\newblock {{\it Science} {\bf 319}, 1220--1223} (2008).

\bibitem[62]{mod+2008}
{{Modjaz}, M., {Kirshner}, R.~P., {Blondin}, S., {Challis}, P. \& {Matheson}, T.} {Double-Peaked Oxygen Lines Are Not Rare in Nebular Spectra of Core-Collapse Supernovae}.
\newblock {{\it Astrophys. J. Lett} {\bf 687}, L9--L12} (2008).

\bibitem[63]{tau+2009}
{{Taubenberger}, S., {Valenti}, S., {Benetti}, S., {Cappellaro}, E., {Della Valle}, M. {\it et al.}} {Nebular emission-line profiles of Type Ib/c supernovae - probing the ejecta asphericity}.
\newblock {{\it Mon. Not. R. Astron. Soc.} {\bf 397}, 677--694} (2009).

\end{thebibliography}
\end{document}